\documentclass{article}

\usepackage{arxiv}

\usepackage[utf8]{inputenc} 
\usepackage[T1]{fontenc}    
\usepackage{url}            
\usepackage{booktabs}       
\usepackage{amsfonts}       
\usepackage{nicefrac}       
\usepackage{microtype}      
\usepackage{lipsum}		
\usepackage{graphicx}
\usepackage{doi}
\usepackage{amsmath}
\usepackage{amsthm}
\usepackage{color}

\newtheorem{theorem}{\bf Theorem}[section]

\usepackage{bm}
\usepackage{lmodern}

\newcommand{\rmi}{{\rm i}} 
\newcommand{\conj}[1]{\overline{#1}}
\renewcommand{\d}{{\rm d}}

\title{Free energy formulas for
conﬁned nematic liquid
crystals based on analogies
with Kirchhoﬀ-Routh theory
in vortex dynamics}

\author{{\hspace{1mm}Hiroyuki Miyoshi$^1$\thanks{Corresponding author: hiroyukimiyoshi@g.ecc.u-tokyo.ac.jp},\ \  Hiroki Miyazako$^{1,2}$,\ \ Takaaki Nara$^{1,2}$} \\ \\ 
	$^1$Graduate School of Information Science and
Technology, \\The University of Tokyo, Hongo,
7-3-1, Bunkyo-Ku, Tokyo, 113-8656, Japan \\
$^2$Department of Information Physics and
Computing, \\The University of Tokyo, Hongo,
7-3-1, Bunkyo-Ku, Tokyo, 113-8656, Japan \\
}

\renewcommand{\shorttitle}{Free energy formulas for nematic liquid crystals}

\hypersetup{
pdftitle={Free_Energy_formula},
pdfsubject={q-bio.NC, q-bio.QM},
pdfauthor={Hiroyuki Miyoshi, Hiroki Miyazako, and Takaaki Nara},
pdfkeywords={topological defects, conformal mappings, nematic liquid crystals},
}

\begin{document}
\maketitle

\begin{abstract}
Active nematics are influenced by alignment angle singularities called topological defects.  
    The localization of these defects is of major interest for biological applications. 
    The total distortion of alignment angles due to defects is evaluated using Frank free energy, 
    which is one of the criteria used to determine the location and stability of these defects.
    Previous work used the line integrals of a complex potential associated with the alignments for the energy calculation (Miyazako and Nara, R. Soc. Open Sci., 2022), which 
    has a high computational cost. 
    We propose analytical formulas 
    for the free energy in the presence of multiple topological defects in confined geometries. 
    The formulas derived here are an analogue of Kirchhoff-Routh functions in vortex dynamics.  
    The proposed formulas are explicit with respect to the defect locations and conformal maps, which enables the explicit calculation of the energy extrema. 
    The formulas are applied to calculate the locations of defects in so-called doublets and triplets by solving simple polynomial formulas.  
    A stability analysis is also conducted to detect whether defect pairs with charges $\pm 1/2$ are stable or unstable in triplet regions. 
    Our numerical results are shown to match the experimental results (Ienaga {\em et al.,} Soft Matter, 2023). 
\end{abstract}


\section{Introduction}
A spindle-shaped cell is characterized by its elongated shape. 
Such cells play crucial roles in various physiological processes, including muscle contraction, tissue organization, and cell migration. 
Investigations of the dynamics of biological cells are important for elucidating fundamental processes in development, disease progression, and tissue homeostasis. 
In such investigations, spindle- and rod-shaped cells are modeled as nematic liquid crystals, which are also known as ``active nematics''~\cite{zhao2019advances,doostmohammadi2022physics}. 
Active nematics are essential not only for understanding the collective dynamics of biological systems such as cells and tissues~\cite{Woltman2007-jl} but also for artificial biological systems. 
The various phenomena observed in active nematics under diverse scenarios have been reviewed in~\cite{marchetti2013hydrodynamics}.

An important property in the field of active nematics 
is the existence of topological defects~\cite{de1993physics,doostmohammadi2022physics}. 
Topological defects are regions where the alignment angles of nematic liquid crystals are not defined. 
Such singularities influence the orientation of surrounding nematics~\cite{aranson2019topological} and determine the physical behaviours of active nematics. 
The information of the rotations in the surrounding alignments 
is referred to as the topological charge of defects, which takes discrete values of integers or half-integers.

Topological defects play a significant role in biological systems. 
Maroudas-Sacks {\em et al}. presented experimental data on topological defects in {\em Hydra} and showed 
that topological defects with topological charges $-1/2$, $+1/2$, and $+1$ regulate the process of morphogenesis~\cite{maroudas2021topological}. 
Kawaguchi {\em et al.} found that topological defects drive the aggregation of cultured murine neural progenitor cells~\cite{Kawaguchi2017-df}.
They demonstrated that cells concentrate around defects with a charge of $+1/2$, and escape from defects with a charge of $-1/2$.

The behaviour of topological defects is influenced by the presence of boundaries. 
Saw {\em et al.} observed topological defects with charges of $+1/2$ and $-1/2$ in a star-shaped region to study cell death and extrusion~\cite{saw2017topological}. 
Ienaga {\em et al}. modified the closed domains of single circular shapes 
to regions consisting of multiple discs overlapping each other (doublets and triplets) to study the relationship between existing defects and geometries~\cite{Ienaga2023-qm}. 
They conjectured that the positions of defects in both doublets and triplets linearly correlate with the distances between the centres of circles. 
Furthermore, they experimentally examined the transition from a formation where two defects with a charge of $+1/2$ exist 
to a state where a defect with a charge of $-1/2$ and three defects with a charge of $+1/2$ exist 
as the three discs of triplets move apart.

Based on liquid crystal theory and one-constant approximation~\cite{de1993physics}, 
mathematical analysis using the theory of two-dimensional (2D) harmonic functions has been conducted to determine the location and stability of topological defects in confined geometries. 
Duclos {\em et al.} investigated the positions of topological defects in circular closed domains 
and confirmed both experimentally and mathematically 
that two defects with a charge of $+1/2$ are located at a distance of $5^{-1/4}R$ 
from the centre of the disc with radius $R$~\cite{Duclos2016-gn}. 
They modelled the Frank free energy associated with cell alignment angles based on one-constant approximation 
and mathematically proved that defects exist at the minimum of this free energy. 
They expressed the energy as the surface integral of the square of the spatial gradients of orientation excluding the vanishingly small defect core regions. 

Because the alignment angles are harmonic functions in 2D, it is natural to use a complex analysis formulation. 
Tarnavskyy {\em et al.} showed that topological defects exist on the edge of a half of a lens-shaped region using conformal mappings~\cite{tarnavskyy2018generalised}. 
Chandler and Spagnolie used Schwarz-Christoffel mappings to find topological defects on the boundary of a polygonal-shaped immersed body~\cite{chandler2023nematic}.
The theory developed by Chandler and Spagnolie was also applied to study two-body interactions such as body forces and torques~\cite{chandler2023exact}. 
Tang {\it et al.} showed that Frank free energy can be classified into torque-free energy and orientation energy and 
that torque determines the dynamics of topological defects~\cite{Tang2017-nb}. 

Miyazako and Nara proposed a complex potential for cell alignment angles 
based on complex analysis to represent the orientational order of cells~\cite{Miyazako2022-sh}. 
They derived an analytical formula that describes 
the orientation of cells with topological defects at arbitrary positions 
strictly inside arbitrary closed domains~\cite{Miyazako2022-sh}. 
They employed the theory of conformal mapping 
to compute cell orientations in arbitrary shapes. 
Additionally, by utilizing the complex Green's theorem, 
they evaluated the energy of the domain as line integrals to avoid the divergence of energy at singularities of defects. 
They updated defect positions using steepest descent methods, 
and discussed whether the initial configuration of defects would annihilate or stabilize in star-shaped domains. 
Recently, this approach was extended to doubly connected domains~\cite{Miyazako_undated-ue} using the prime functions developed by Crowdy~\cite{crowdy2020solving_book}.

Following Duclos {\it et al.}~\cite{Duclos2016-gn}, we assume that topological defects become stable when the Frank free energy is minimized. 
Energy calculations are necessary for localizing topological defects at arbitrary positions in confined domains. 
However, there are no analytical formulas for the Frank free energy in these geometries except for the line integrals proposed by Miyazako and Nara~\cite{Miyazako2022-sh}, 
which have a high numerical cost and require finding a suitable branch cut of the logarithmic singularities associated with defects. 
This significantly complicates the theoretical analysis of the positions of topological defects.

\begin{figure}[t]
    \centering
    \includegraphics[bb=200 0 500 300,scale=0.53]{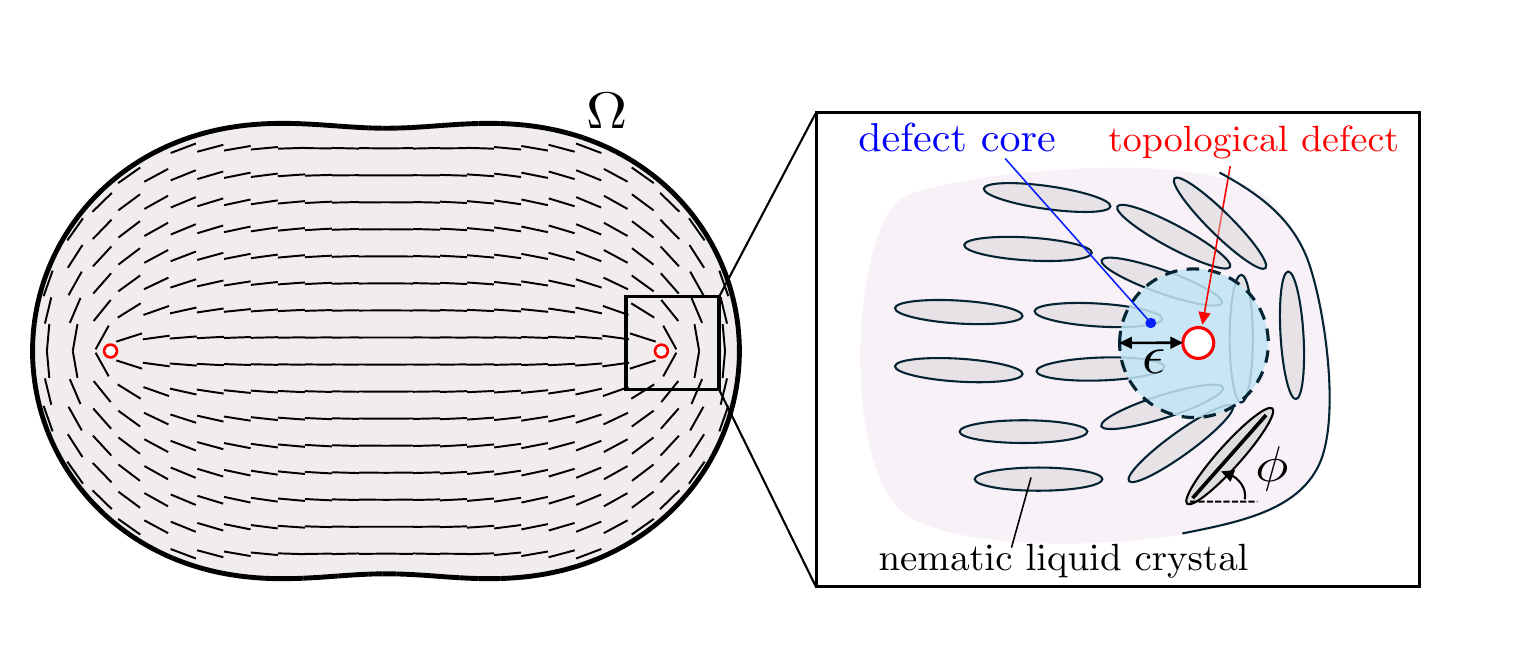}
    \caption{Schematic diagram of topological defects (red circles) with charge of $+1/2$ and resulting alignments due to defects and boundary. 
    The alignment angle changes by $\pi$ around the topological defects. 
    The defects here are assumed to have a defect core with size $\epsilon$, 
    which is sufficiently small compared to the outer boundary of confined domain $\Omega$. }
    \label{fig:nematic}
\end{figure}

In this paper, we develop analytical formulas for the Frank free energy when the radius of the defect core is small as shown in Figure~\ref{fig:nematic}. 
The explicit formulas are derived from an expression for the 
complex potential, representing the sum of each defect, excluding the contribution from the defect itself. 
The proposed formulas are new in the field of liquid crystal theory. 
This paper provides an analogy between the concept of Hamiltonian or Kirchhoff-Routh functions for vortex dynamics~\cite{lin1941motion,crowdy2020solving_book} and the Frank free energy of nematic liquid crystals. 
The proposed formulas enable to derive the explicit expression for the energy up to the order
$\epsilon\log\epsilon$, where $\epsilon$ is the size of defect cores. 
It is mathematically demonstrated that
two defects are located symmetrically at $\pm 5^{-1/4}$ in the unit disc. 
We also validate the conjecture regarding the positions of topological defects in doublets and triplets given by Ienaga {\em et al.}~\cite{Ienaga2023-qm}. 
It is found that our analytical results closely match the experimental results for doublets reported by Ienaga {\em et al.}, 
and detect the transition of the formation of topological defects in triplets.

The rest of this paper is organized as follows. 
Section~\ref{sec:math_formulation} introduces a mathematical formulation of alignment with topological defects 
and the energy of alignment angles. 
Section~\ref{sec:3} presents the proposed formulas for the Frank free energy in the unit disc and arbitrarily shaped regions. 
Section~\ref{sec:4} calculates the locations of topological defects in doublets and triplets based on the theory of quadrature domains~\cite{gustafsson2005quadrature} and 
the stability of these defects. 
Section~\ref{sec:5} gives the conclusions and provides future prospects.

\section{Formulation of topological defects and Frank-free energy}
\label{sec:math_formulation}
\subsection{Problem setting}
Following the theory developed by Miyazako and Nara~\cite{Duclos2016-gn}, 
this section introduces a mathematical formulation of alignment angles in the presence of multiple topological defects based on complex analysis. 
The nematic liquid crystals are assumed to be spindle-shaped and small relative to the outer boundary, thus allowing them to be treated as rod-like particles. 
They are assumed to be aligned in the direction of a unit vector $\bm{n}_{\rm d}$ as follows:
\begin{align}
    \bm{n}_{\rm d} = (\cos\phi(u,v),\sin \phi(u,v))^\top,
\end{align}
where $\phi(u,v)$ is the angle of the field at position $(u,v)$. 
The direction $\bm{n}_{\rm d}$ is called a ``director''. 
Since $\bm{n}_{\rm d}$ and $-\bm{n}_{\rm d}$ are indistinguishable, an arbitrary angle $\phi$ is equivalent to $\hat{\phi}\in [-\pi/2,\pi/2)$~\cite{Miyazako2022-sh}; that is, 
the two angles $\phi_1$ and $\phi_2$ are equivalent if there exists an integer $n\in \mathbb{Z}$ such that $\phi_1 = \phi_2 + n\pi$. 

The governing equation for the alignment angle $\phi$ is derived using the one-constant approximation of elastic energy called Frank free energy and its variational formula. 
The Frank free energy with $\phi(u, v)$ at position $(u, v)$ 
is defined as the integral of a surface $\Omega$ as follows:
\begin{align}
    F \equiv \frac{K}{2}\int_{\Omega}|\nabla \phi(u, v)|^2 \d u \d v,\label{eq:Frank}
\end{align}
where $K$ is a Frank elastic constant and $K/2=1$ in this paper for simplicity. 
The variation of the energy defined in equation~(\ref{eq:Frank}) leads to 
the energy taking a minimum when 
the alignment angle $\phi(u, v)$ satisfies the following 2D Laplace's equation:
\begin{align}
    \nabla^2 \phi(u, v) = 0,\quad (u, v)\in \Omega. 
\end{align}

\begin{figure}[t]
    \centering
    \includegraphics[bb=200 0 1100 850,scale=0.34]{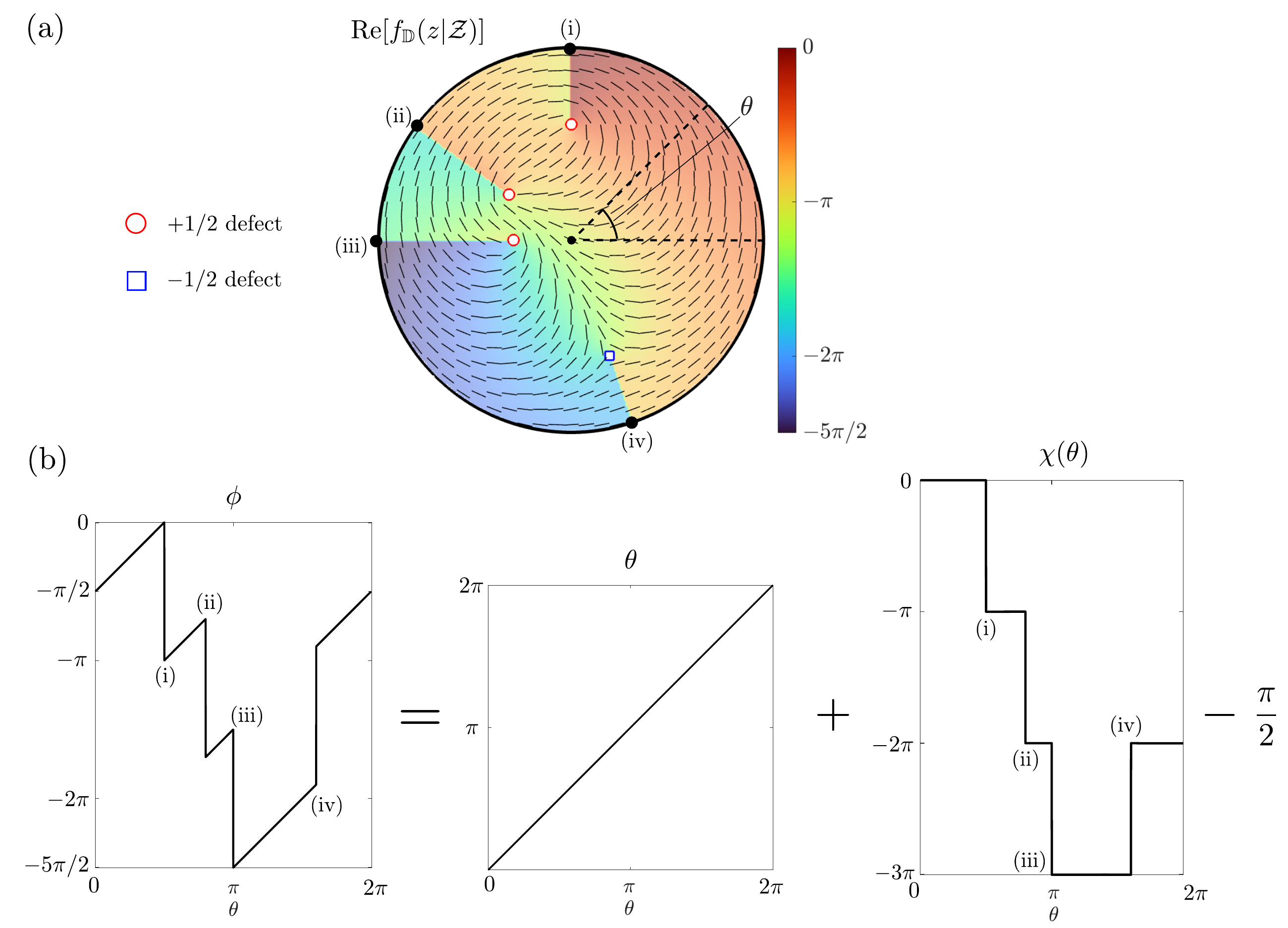}
    \caption{(a) Colour plot of ${\rm Re}[f_\mathbb{D}(z|\mathcal{Z})]$ with three defects with charge of $+1/2$, and one defect with charge of $-1/2$. 
    Four branch cuts exist due to the logarithmic singularities in the unit disc. 
    (b) Angle $\phi$ on the unit circle decomposed into $\theta$ and step function $\chi(\theta)$ associated with branch cuts minus $\pi/2$. 
    Note that $\theta+\chi(\theta)$ on $\partial \mathbb{D}$ is discontinous at (i)-(iv). }
    \label{fig:abst}
\end{figure}

A complex analysis formulation of the alignment for simply connected domains is now introduced. 
First, an analytical formula for the unit disc is introduced. 
Let $z=x+\rmi y$ define a complex plane and $\mathbb{D}$ define the unit disc. 
There exist topological defects at $z_k=x_k + \rmi y_k$, $k=1,\ldots,N$, strictly inside $\mathbb{D}$, whose topological charges are $q_k$. 
The total number of topological charges is $N$ and the charges are integers or half-integers except $0$; that is: 
\begin{align}
    q_k \in \{n/2 | \ n\in \mathbb{Z},\ n\neq 0\}.
\end{align}
We denote the set of positions of defects as $\mathcal{Z} \equiv \{z_k|k=1,2,\ldots,N\}$. 

The alignment angle $\phi(x,y)$ created by these topological defects satisfies the following conditions~\cite{Duclos2016-gn,Miyazako2022-sh}:
\begin{enumerate}
    \item Alignment angle $\phi$ satisfies 2D Laplace's equation except at $\mathcal{Z}$:
    \begin{align}
        \nabla^2 \phi(x,y) = 0,\quad (x,y)\in \mathbb{D}\backslash \mathcal{Z}.
    \end{align}
    \item The sum of topological charges $q_k$ is equal to $1$~\cite{Miyazako2022-sh}; that is:
    \begin{align}
        \sum_{k=1}^N q_k = 1.  \label{eq:cond_sumq}
    \end{align}
    \item For simply connected domains $D_k\subset \mathbb{D}$ $(k=1,\ldots,N)$ that contain only the $k$-th defect, the circulation around $D_k$ satisfies 
    \begin{align}
        \oint_{\partial D_k} \frac{\partial \phi}{\partial s} \d s = 2\pi q_k.
    \end{align}
    \item Because the directors align along the boundary of $\mathbb{D}$, alignment angle $\phi(x, y)$ satisfies 
    \begin{align}
        \phi(x,y) = \nu(x,y),\quad (x,y)\in \partial \mathbb{D},
    \end{align} 
    where $\nu(x,y)$ is a tangential angle on $\partial \mathbb{D}$. 
\end{enumerate}

Miyazako and Nara derived an analytical formula that describes the alignment angle that satisfies conditions (i)-(iv) in $\mathbb{D}$~\cite{Miyazako2022-sh}. 
The alignment angle $\phi(x,y)$ for the unit disc is given by 
$\phi(x,y) = {\rm Re}[f_\mathbb{D}(z|\mathcal{Z})]$, where $f_\mathbb{D}(z|\mathcal{Z})$ is an analytic function in domain $\mathbb{D}\backslash\mathcal{Z}$ which is explicitly given by 
\begin{align}
    f_\mathbb{D}(z|\mathcal{Z}) = \sum_{k=1}^N q_k \mathcal{G}(z,z_k) - \frac{\pi}{2}, \label{eq:fD}
\end{align}
where a complex function for the cell alignment angles is defined by 
\begin{align}
    \mathcal{G}(z,z_k) \equiv -\rmi (\log(z-z_k) + \log(1-\conj{z_k}z)),\label{eq:Green}
\end{align}
where $\conj{a}$ represents the complex conjugate of complex number $a$. Topological charge $q_k$ satisfies condition~(\ref{eq:cond_sumq}). 
Figure~\ref{fig:abst} (a) shows an example of $\phi$ in the presence of three topological defects with a charge of $+1/2$ and a defect with a charge of $-1/2$.

Miyazako and Nara extended the analytical formula to simply connected arbitrarily shaped domains using conformal mappings. 
Riemann's mapping theorem states that there exists a conformal mapping $w=g(z)$ 
from the unit disc $\mathbb{D}$ in the complex $z$-plane to the region $\Omega$ in a complex $w=u+\rmi v$-plane.
The positions of topological defects inside $\Omega$ are denoted as $w_k$, $k=1,\ldots,N$, and the set of these positions is $\mathcal{W}\equiv \{w_1,w_2,\ldots,w_N\}$. 
The preimage of $w_k$ is denoted as $z_k$; that is, $z_k = g^{-1}(w_k)\in \mathbb{D}$, $z_k\notin \partial \mathbb{D}$. 
Using the map $w=g(z)$, the tangential angle on the boundary of $\Omega$ is given by 
\begin{align}
    \nu(\theta) = {\rm Re}\left[-\rmi \log \left(\frac{du}{d\theta} + \rmi \frac{dv}{d\theta} \right) \right]= {\rm Re}\left[-\rmi \log g'(z) +\theta + \frac{\pi}{2} \right],\quad z=e^{\rmi\theta},
\end{align}
which gives the following complex potential of cell alignment for $\Omega$~\cite{Miyazako2022-sh,Miyazako_undated-ue}: 
\begin{align}
    f_\Omega(z|\mathcal{Z}) = f_\mathbb{D}(z|\mathcal{Z}) - \rmi \log g'(z),\quad w = g(z). 
\end{align}

\subsection{Line integral for energy calculations}
\begin{figure}[t]
    \centering
    \includegraphics[bb=-100 0 700 450,scale=0.40]{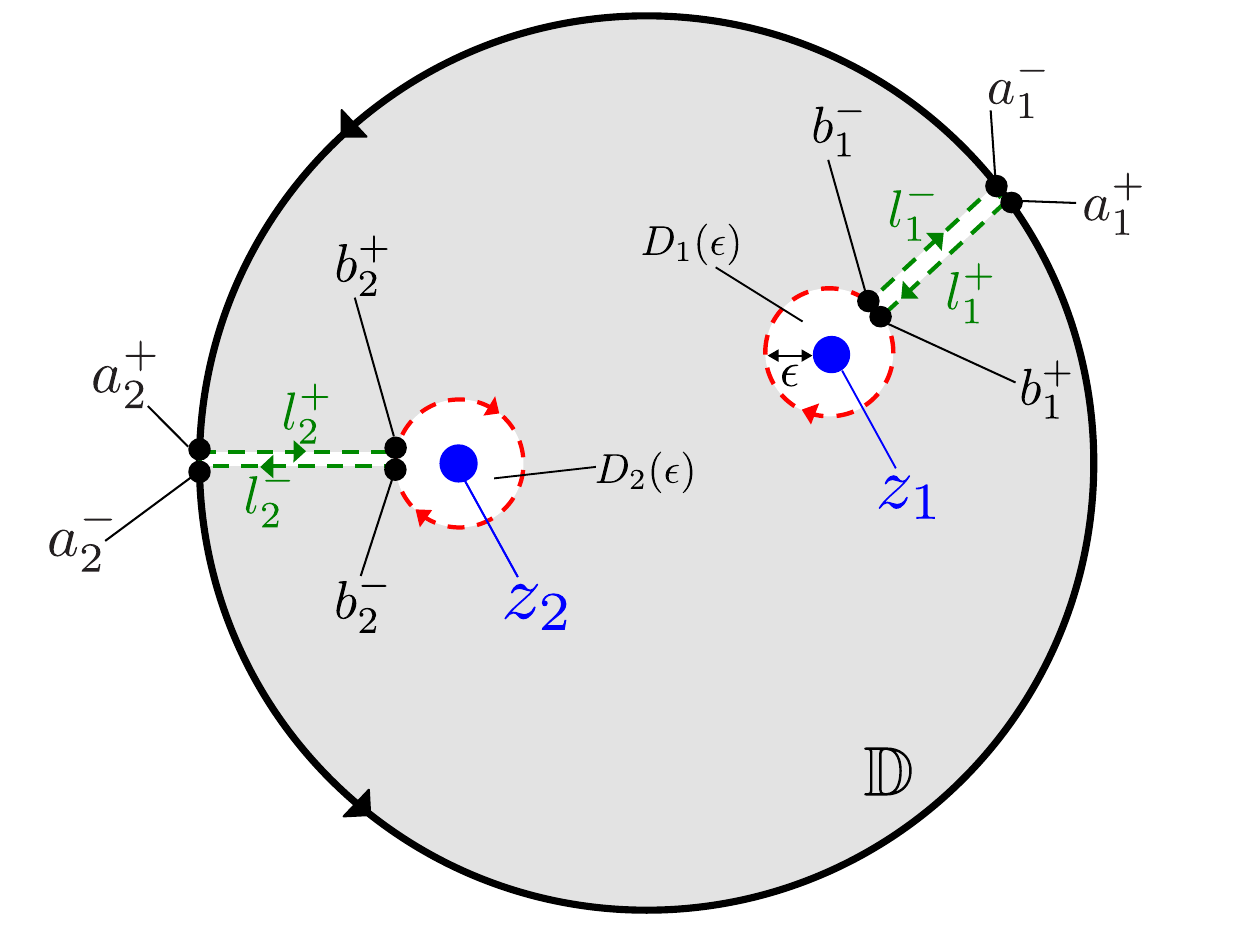}
    \caption{Line integral along $\partial (\mathbb{D}\backslash D(\epsilon))$ for case $N=2$. Branch cuts of logarithmic singularities $z_k$ are chosen as $l_k$ and the imaginary part of $\log(z-z_k)$ jumps on $l_k$ by $2\pi$. }
    \label{fig:lineint_circle}
\end{figure}

The positions of topological defects $w_k$ tend to be located where the energy defined by equation~(\ref{eq:Frank}) satisfies the local minimum~\cite{Duclos2016-gn,Miyazako2022-sh}. 
Due to the presence of topological defects at $w_k$, the surface integral of $|\nabla \phi|^2$ is divergent and thus 
special care must be taken. 
Duclos {\em et al}.~\cite{Duclos2016-gn} used the Fourier expansions and 
calculated the energy in the disc with radius $R$ except for small regions around the defects. 
They showed that the free energy of the disc is minimized when there exist two defects located in the disc at a distance $\pm 5^{-1/4}R$ from the origin. 
Miyazako and Nara modified the calculations while preventing the energy divergence of $|\nabla \phi|^2$ at the defects and calculated the energy in star-shaped domains~\cite{Miyazako2022-sh}. 
With the use of the complex Green's theorem, the energy~(\ref{eq:Frank}) is written as the following line integral: 
\begin{align}
    F = \int_{\Omega(\epsilon)}|\nabla \phi(u,v)|^2 \d u \d v = \frac{1}{2\rmi}\oint_{\partial \Omega(\epsilon)} \frac{\partial f_\Omega}{\partial w}\conj{f_\Omega}dw,\quad \frac{\partial }{\partial w}\equiv\frac{1}{2}\left(\frac{\partial}{\partial u} - \rmi \frac{\partial }{\partial v}\right), \label{eq:line_int}
\end{align}
where a small disc around the $k$-th defect with radius $\epsilon$ is defined by 
\begin{align}
    D_k(\epsilon)\equiv \{w_k+re^{\rmi \theta}|\ 0\leq r\leq \epsilon, 0\leq \theta < 2\pi\},
\end{align}
and $\Omega(\epsilon)\equiv \Omega\backslash \cup_{k=1}^N D_k$. 
The path of the line integral for the case $N=2$ is illustrated in Figure~\ref{fig:lineint_circle}. 

Using the line integral formula~(\ref{eq:line_int}), 
Miyazako and Nara confirmed that for the unit disc, 
the energy is minimized when defects exist at positions $z=\pm 5^{-1/4}$. 
They verified that when there are multiple $+1/2$ defects and $-1/2$ defects within the unit disc, 
the pair of positive and negative defects annihilate through energy minimization. 
They also showed that this annihilation happens on symmetrical domains that contain multiple convex regions such as triplets. 

Although the line integral~(\ref{eq:line_int}) is evaluated in a simple manner, 
the logarithmic branch cuts of $f_\Omega$ exist between $\partial D_k(\epsilon)$ and $\partial \Omega$.
When there exist multiple defects, 
special care must be taken when selecting the paths to prevent these branch cuts from crossing.

\section{Analytical formulas for the Frank free energy} \label{sec:3}
This section proposes an analytical formula for the energy defined by~(\ref{eq:Frank}) in the unit disc $\mathbb{D}$ and in arbitrary domain $\Omega$. 
This can be done by splitting equation~(\ref{eq:fD}) into 
a singular part that includes a logarithmic pole at $z=z_k$ and 
a regular part around $z=z_k$. 
The concept of this function is identical to that of the Kirchhoff-Routh functions in the field of vortex dynamics~\cite{lin1941motion,saffman1995vortex}. 
This enables the explicit evaluation of the line integral (\ref{eq:line_int}), 
which gives an analytical formula for the energy up to the order $\mathcal{O}(\epsilon \log \epsilon)$.

\subsection{Analytical formula for unit disc}
\begin{theorem}
    Assume that there exist $N$ topological defects with an integer or half-integer charge located at $z_k\in \mathbb{D}$, $k=1,\ldots,N$, $z_k\notin \partial \mathbb{D}$. 
    The sum of the topological charges satisfies~(\ref{eq:cond_sumq}). 
    Then the total energy $F$ of the region $\mathbb{D}\backslash D(\epsilon)$, $D(\epsilon)\equiv \cup_{k=1}^N D_k(\epsilon)$ is written as follows: 
    \begin{align}
        F = -2\pi \sum_{k=1}^N q_k^2 \log \epsilon + 2\pi F_d + \mathcal{O}(\epsilon \log \epsilon), \label{eq:explicit_unit}
    \end{align}
    where the first term is independent of $z_k$ and 
    $F_d$ only depends on the position of the defects $z_k$, and is given by 
    \begin{align}
        &F_d = \sum_{k=1}^N q_k^2 {\rm Im}[\mathcal{R}(z_k)] + \sum_{k=1}^N\sum_{l\neq k}^N q_k q_l {\rm Im}[\mathcal{G}(z_k,z_l)],
    \end{align}
    where $\mathcal{G}(.,.)$ is given in equation~(\ref{eq:Green}) and 
    \begin{align}
        \mathcal{R}(z_k) = \lim_{z\rightarrow z_k}\left(\mathcal{G}(z,z_k) + \rmi \log (z-z_k)\right) = -\rmi \log(1-z_k\conj{z_k}) - \frac{\pi}{2}.
    \end{align}
    The function $\mathcal{G}(z)+\rmi\log(z-z_k)$ is a function analytic at $z=z_k$, which is obtained by 
    removing the contribution of the logarithmic pole at $z=z_k$. 
\end{theorem}

{\bf Remark 1:}
The first term in (\ref{eq:explicit_unit}) is independent of locations $z_k$, which means that $z_k$ only contributes the second term in (\ref{eq:explicit_unit}). 
The minimum of the energy is obtained by minimizing $F_d$. 

\vspace{5pt}

{\bf Remark 2:}
When we assume that two topological defects with a charge of $q_1=q_2=+1/2$ are located at $z=\pm r$, $0<r<1$, 
the energy $F_d$ for this case is given by 
\begin{align}
    F_d =  -\frac{1}{2}[\log(1-r^2) + \log(1+r^2) + \log(2r)].
\end{align}
Taking derivative with respect to $r$ gives 
\begin{align}
    \frac{\partial F_d}{\partial r} = \frac{5r^4- 1}{2r(1-r^4)}.
\end{align}
Solving the equation $\partial F_d/\partial r=0$, we have $r=5^{-1/4}$, which is consistent with the results given by Duclos {\em et al.}~\cite{Duclos2016-gn}. 

\vspace{5pt}

{\bf Remark 3:}
Our results show that the concept of this expression is identical to that of Hamiltonians for multiple vortices. 
In classical fluid dynamics, Hamiltonians are referred to as Kirchhoff-Routh path functions, which give the vortex trajectories (see p.\ 121 in~\cite{newton2002n}). 
The theory of Kirchhoff-Routh path functions was extended to multiply connected domains by Lin~\cite{lin1941motion} 
and a mathematical formulation for multiply connected domains with prime functions was given in a pioneering work by Crowdy~\cite{crowdy2005analytical}. 
The concept presented here has been used to derive a variational formulation for vortex dynamics 
based on the principle of least action and asymptotic formulas for the kinematic energy have been obtained~\cite{khalifa2024vortex}. 
A Hamiltonian $\mathcal{H}$ in the presence of multiple point vortices with circulation $\Gamma_k$, $k=1,\ldots,N$ is given by~\cite{newton2002n}
\begin{align}
    \mathcal{H} =  - \frac{1}{2}\sum_{k=1}^N\Gamma_k^2 R(z_k) -\frac{1}{2}\sum_{k=1}^N\sum_{k\neq l}^N\Gamma_k \Gamma_l G(z_k,z_l),
\end{align}
where $G(.,.)$ is a hydrodynamic Green's function for a point vortex and $R(.)$ is a Robin function~\cite{flucher1999vortex} for the Green's function $G(.,.)$ and is defined by 
\begin{align}
    R(z_k) = \lim_{z\rightarrow z_k}\left(G(z,z_k) - \frac{1}{2\pi}\log|z-z_k|\right).
\end{align}
Note that the Green's function for a point vortex in the unit disc is given by 
\begin{align}
    G(z,z_k) \equiv \frac{1}{2\pi}\log\left|\frac{z-z_k}{|z_k|(z-1/\conj{z_k})} \right|.
\end{align}
The imaginary part of $\mathcal{G}(z,z_k)$ is similar to $-2\pi G(z,z_k)$ except for the opposite sign of the second term in~(\ref{eq:Green}) due to the boundary conditions. 
Our result indicates that the classical approach used in vortex dynamics can be extended to the theory of nematic liquid crystals.

\begin{proof}
For simplicity, we define 
\begin{align}
    r_k(z) \equiv f_\mathbb{D}(z|\mathcal{Z}) + \rmi q_k \log (z-z_k). \label{eq:rk}
\end{align}
This function is analytic at $z=z_k$ and it is easy to see that $r_k(z)$ satisfies 
\begin{align}
    \lim_{z\rightarrow z_k}r_k(z) = q_k\mathcal{R}(z_k) +  \sum_{l\neq k}^N q_l \mathcal{G}(z_k,z_l) - \frac{\pi}{2}.  \label{eq:robin-type}
\end{align}
The branch cuts of $\log(z-z_k)$
are chosen such that they do not intersect each other. 
The path of the line integral is shown in Figure~\ref{fig:lineint_circle}. 
In the figure, $l_k^+$, $l_k^-$ denote the paths of the line integral from $\partial \mathbb{D}$ to $\partial D_k(\epsilon)$ used to avoid the jumps of the branch cuts of $\log(z-z_k)$. 
It is noted that for $z^\pm\in l_k^\pm$, the following property is always satisfied:
\begin{align}
    \log (z^+-z_k) - \log(z^- - z_k) = 2\pi\rmi.
\end{align}
Let $a_k^+$ and $a_k^-$ denote the intersections of the paths $l_k^+$ and $l_k^-$ with $\partial \mathbb{D}$, and 
$b_k^+$ and $b_k^-$ denote the intersections of $l_k^+$ and $l_k^-$ with $\partial D_k(\epsilon)$, respectively. 
We choose $0\leq {\rm arg}(a_1^{\pm})<{\rm arg}(a_2^{\pm})<\ldots<{\rm arg}(a_N^{\pm})<2\pi$ as shown in Figure~\ref{fig:abst}. 
Due to the existence of branch cuts, angle $\phi$ has jumps at these points. 
We define the argument on the boundary $\theta$ as shown in Figure~\ref{fig:abst} (a). 
For $z=e^{\rmi\theta}$, $\theta\in[0,2\pi)$, 
\begin{align}
    {\rm Re}[f(e^{\rmi\theta}|\mathcal{Z})] = \phi(\theta) =\theta + \chi(\theta) - \frac{\pi}{2}, \label{eq:phase_jump}
\end{align}
where $\chi(\theta)$ is a step function whose value jumps at $a_k^{\pm}$ as follows:
\begin{alignat}{2}
    \chi(\theta) = \left\{\begin{aligned}
        &0,\quad &&{\rm for}\  \theta\in [0,{\rm arg}(a^+_{1})],\\
        &-\sum_{k=1}^{l} 2\pi q_k \quad &&{\rm for}\  \theta\in [{\rm arg}(a_l^-),{\rm arg}(a^+_{l+1})],\quad l = 1,\ldots, N-1,\\
        &-2\pi,\quad &&{\rm for}\ \theta \in [{\rm arg}(a^-_{N}),2\pi).
    \end{aligned}\right.
\end{alignat}
This decomposition is illustrated in Figure~\ref{fig:abst} (b). 

Using the complex Green's theorem,  the line integral of the energy~(\ref{eq:line_int}) is given by 
\begin{align}
    F &= \frac{1}{2\rmi}\left[\sum_{k=1}^N \int_{a_k^-}^{a_{k+1}^+} \frac{\partial f_\mathbb{D}}{\partial z}\conj{f_\mathbb{D}} \d z + \sum_{k=1}^N \left(\int_{l_k^+} \frac{\partial f_\mathbb{D}}{\partial z}\conj{f_\mathbb{D}} \d z -  \int_{l_k^-} \frac{\partial f_\mathbb{D}}{\partial z}\conj{f_\mathbb{D}} \d z - \oint_{\partial D_k(\epsilon) } \frac{\partial f_\mathbb{D}}{\partial z}\conj{f_\mathbb{D}} \d z \right)\right], \label{eq:line_intC0_all}
\end{align}
where $a_{N+1}^{+} = a_1^+$. 
We evaluate each term in equation~(\ref{eq:line_intC0_all}) using the functions defined by~(\ref{eq:rk}). 

\vspace{5pt}

\textbf{(i) Line integral along $\partial D_k(\epsilon)$}

The arguments of the function $\log(z-z_k)$ at $z=b_k^\pm$ are set to ${\rm arg}(b_k^{\pm}-z_k) = \beta_k^\pm$. 
Using the change of variables along $\partial D_k(\epsilon)$ such that $z=z_k + \epsilon e^{\rmi\theta}$, $\theta\in[\beta^-_k ,\beta_k^+]$ and equation~(\ref{eq:rk}), the line integral of $\partial D_k(\epsilon)$ is evaluated as follows:
\begin{align}
    -\frac{1}{2\rmi}\oint_{\partial D_k(\epsilon)} \frac{\partial f_\mathbb{D}}{\partial z} \conj{f_\mathbb{D}} \d z &= -\frac{1}{2\rmi}\int_{\beta_k^-}^{\beta_k^+} \left[-\frac{\rmi q_k}{\epsilon e^{\rmi\theta}} + \frac{\partial r_k}{\partial z}(z_k + \epsilon e^{\rmi\theta}) \right]\left[\rmi q_k\log (\epsilon e^{-\rmi\theta}) + \conj{r_k(z_k+\epsilon e^{\rmi\theta})} \right]\rmi \epsilon e^{\rmi \theta} \d\theta\nonumber\\
    &=-\frac{1}{2}\int_{\beta_k^-}^{\beta_k^+}\left[-\rmi q_k \conj{r_k(z_k)} + q_k^2 \log \epsilon - \rmi q_k^2 \theta \right]  \d \theta  + \mathcal{O}(\epsilon\log\epsilon)\nonumber \\ 
    &=-\pi q_k^2 \log \epsilon + \rmi \pi q_k \conj{r_k(z_k)} + \rmi \pi^2 q_k^2 + \rmi \pi  q_k^2 \beta_k^-  + \mathcal{O}(\epsilon\log\epsilon) \label{eq:int_Dk}.
\end{align}
where we used that $r_k(z_k+\epsilon e^{\rmi\theta})= r_k(z_k) + \mathcal{O}(\epsilon)$ and $\beta_k^+ - \beta_k^- = 2\pi$ from the second to third line. 

\vspace{5pt}

\textbf{(ii) Integrals along $l_k^+$ and $l_k^-$}

The line integrals along $l_k^{\pm}$ are given by 
\begin{align}
\frac{1}{2\rmi}\int_{l_k^\pm}\frac{\partial f_\mathbb{D}}{\partial z} \conj{f_\mathbb{D}} \d z = \frac{1}{2\rmi}\int_{a_k^\pm }^{b_k^\pm} \left(-\frac{\rmi q_k}{z-z_k}+ \frac{\partial r_k}{\partial z} \right) \left(\rmi q_k\log\conj{(z^{\pm}-z_k)} + \conj{r_k(z)} \right) \d z,
\end{align}
where $z^{\pm}$ are points on $l_{k}^{\pm}$. 
The line integral along $l_k^+$ minus the integral along $l_k^-$ becomes 
\begin{align}
    \frac{1}{2\rmi}\int_{l_k^+}\frac{\partial f_\mathbb{D}}{\partial z} \conj{f_\mathbb{D}} \d z &- \frac{1}{2\rmi}\int_{l_k^-}\frac{\partial f_\mathbb{D}}{\partial z} \conj{f_\mathbb{D}} \d z = 2\pi q_k\cdot \frac{1}{2\rmi}\int_{a_k^-}^{b_k^-}\left[-\frac{\rmi q_k}{z-z_k}+ \frac{\partial r_k}{\partial z} \right]\d z\nonumber\\
    &= -\rmi \pi q_k\left[-\rmi q_k (\log(b_k^- - z_k) - \log(a_k^- - z_k)) + r_k(b_k^-) - r_k(a_k^-)\right]. \label{eq:int_lk}
\end{align}
Due to the fact that $\log(b_k^- - z_k) = \log \epsilon + \rmi\beta_k^-$ and $r_k(b_k^-) = r_k(z_k) + \mathcal{O}(\epsilon)$ as $\epsilon\rightarrow 0$, equation~(\ref{eq:int_lk}) becomes 
\begin{align}
    \frac{1}{2\rmi}\int_{l_k^+}\frac{\partial f_\mathbb{D}}{\partial z} \conj{f_\mathbb{D}} \d z &- \frac{1}{2\rmi}\int_{l_k^-}\frac{\partial f_\mathbb{D}}{\partial z} \conj{f_\mathbb{D}} \d z 
    =-\pi q_k^2 \log \epsilon - \rmi\pi q_k^2 \beta_k^- -  \rmi\pi q_k r_k(z_k) + \rmi\pi q_k f_\mathbb{D}(a_k^-|\mathcal{Z}) + \mathcal{O}(\epsilon). \label{eq:sum_akbk}
\end{align}

\vspace{5pt}

\textbf{(iii) Line integral along the unit circle}

First we note that $\partial f_\mathbb{D}/\partial z$ is analytic everywhere in $\mathbb{D}\backslash\mathcal{Z}$. 
Using equation~(\ref{eq:phase_jump}), the line integral of the unit circle becomes 
\begin{align}
    \frac{1}{2\rmi}\sum_{k=1}^N\int_{a_k^-}^{a_{k+1}^{+}}\frac{\partial f_\mathbb{D}}{\partial z} \conj{f_\mathbb{D}} \d z= \frac{1}{2\rmi}\sum_{k=1}^N\int_{a_k^-}^{a_{k+1}^{+}}\frac{\partial f_\mathbb{D}}{\partial z} (2\theta + 2\chi(\theta) -\pi - f_\mathbb{D})\d z.\label{eq:aroundC0}
\end{align}
Let the argument of $a_k^{\pm}$ define $\alpha_k^\pm$. 
Note that $\chi(\alpha_k^+) - \chi(\alpha_k^-)=-2\pi q_k$. 
The first two terms in (\ref{eq:aroundC0}) are 
\begin{align}
    &\frac{1}{\rmi}\sum_{k=1}^N \int_{a_k^-}^{a_{k+1}^+}(\theta + \chi(\theta))\frac{\partial f_\mathbb{D}}{\partial z}  \d z=-\rmi \sum_{k=1}^{N}[(\theta+\chi(\theta)) f_\mathbb{D}(e^{\rmi\theta}|\mathcal{Z})]_{\theta=\alpha_k^-}^{\theta=\alpha_{k+1}^+} + \rmi \sum_{k=1}^N\int_{\alpha_k^-}^{\alpha_{k+1}^+} f_\mathbb{D}(e^{\rmi\theta}|\mathcal{Z})\d\theta \label{eq:firstterm}\\
    &=-\rmi\sum_{k=1}^N(2\pi q_k f_\mathbb{D}(a_k^{-}|\mathcal{Z}) + 4\pi^2 q_k^2 + 2\pi q_k (\chi(\alpha_k^-)+\alpha_k^+)) + \rmi \sum_{k=1}^N (2 \pi^2  q_k^2 + 2 \pi q_k (\chi(\alpha_k^-)+\alpha_k^+) )- \rmi\pi^2 \nonumber \\
    &=-\rmi \sum_{k=1}^N (2\pi q_k f_\mathbb{D}(a_k^{-}|\mathcal{Z}) + 2\pi^2 q_k^2) - \rmi\pi^2,
 \end{align}
where the second term in (\ref{eq:firstterm}) is calculated as
\begin{align}
    \rmi \sum_{k=1}^N \int_{\alpha_k^-}^{\alpha_{k+1}^+}f_\mathbb{D}(e^{\rmi\theta}|\mathcal{Z}) \d\theta &= \rmi \sum_{k=1}^N \int_{\alpha_k^-}^{\alpha_{k+1}^+}{\rm Re}[f_\mathbb{D}(e^{\rmi\theta}|\mathcal{Z})] \d\theta =-\rmi\sum_{k=1}^{N}\int_{\alpha_k^-}^{\alpha_{k+1}^+}\left(\theta + \chi(\theta)-\frac{\pi}{2}\right)\d \theta. 
    \label{eq:complicated}
\end{align}
For the first equality in (\ref{eq:complicated}), we used the fact that when $|t|<1$, 
\begin{align}
    \int_{0}^{2\pi}\log|e^{\rmi\theta} - t| d\theta = \int_{0}^{2\pi}\log|1- \conj{t}e^{\rmi\theta} | d\theta = 0,
\end{align}
which can be proven by the residue theorem for $|t|<1$:
\begin{align}
    \oint_{\partial \mathbb{D}} \frac{\log(1-t z)}{z}\d z = 0.
\end{align}


The third term in (\ref{eq:aroundC0}) is 
\begin{align}
    -\frac{\pi}{2\rmi} \sum_{k=1}^N \int_{a_k^-}^{a_{k+1}^+} \frac{\partial f_{\mathbb{D}}}{\partial z} \d z = -\frac{\pi}{2\rmi} \sum_{k=1}^{N}\left[f_\mathbb{D}(z|\mathcal{Z})\right]_{z=a_k^{-}}^{z=a_{k+1}^+}= \rmi\pi^2  \sum_{k=1}^N q_k  = \rmi \pi^2 .
\end{align}
where we used the fact that $f_\mathbb{D}$ shifts $2\pi q_k$ at $z=a_k^\pm$. 
The fourth term in (\ref{eq:aroundC0}) is 
\begin{align}
    -\frac{1}{2\rmi} \sum_{k=1}^N \int_{a_k^-}^{a_{k+1}^+} \frac{\partial f_{\mathbb{D}}}{\partial z} f_{\mathbb{D}} \d z &= -\frac{1}{4\rmi} \sum_{k=1}^{N}\left[f^2_\mathbb{D}(z|\mathcal{Z})\right]_{z=a_k^{-}}^{z=a_{k+1}^+}= -\frac{1}{4\rmi}\sum_{k=1}^N 2\pi q_k (2f_{\mathbb{D}}(a_k^-|\mathcal{Z}) + 2\pi q_k). \nonumber
\end{align}
Hence, the line integral around $\partial \mathbb{D}$ in~(\ref{eq:aroundC0}) is given by
\begin{align}
    \frac{1}{2\rmi}\sum_{k=1}^N\int_{a_k^-}^{a_{k+1}^{+}}\frac{\partial f_\mathbb{D}}{\partial z} \conj{f_\mathbb{D}} \d z &= -\rmi\pi\sum_{k=1}^N (q_k f_\mathbb{D}(a_k^-|\mathcal{Z}) + \pi q_k^2) \label{eq:around_C0final}.
\end{align}
Combining (\ref{eq:int_Dk}), (\ref{eq:sum_akbk}), and (\ref{eq:around_C0final}), we have 
\begin{align}
    F = -2\pi \sum_{k=1}^N q_k^2 \log\epsilon + 2\pi F_d +\mathcal{O}(\epsilon\log\epsilon),\quad F_d = \sum_{k=1}^Nq_k{\rm Im}[r_k(z_k)].
\end{align}
Using the formula~(\ref{eq:robin-type}), the final expression~(\ref{eq:explicit_unit}) is derived. 
\end{proof}

\subsection{Analytical formula for arbitrarily shaped domain}
The theory of energy calculation on the unit disc shown in the previous section is utilized to derive an explicit formula for energy in arbitrarily shaped regions. 
\begin{theorem}
    We consider an arbitrarily shaped simply connected domain $\Omega$ in the $w$-plane mapped from the unit disc $\mathbb{D}$ by a conformal mapping represented as $w=g(z)$, 
    where the boundary of the domain is sufficiently smooth and $g'(z)\neq 0$ for $z\in \mathbb{D}$.  
    Assume that there exist $N$ topological defects with an integer or half-integer charge located at $w_k\in \Omega$, $k=1,\ldots,N$, $w_k\notin \partial \Omega$, and 
    define the set of topological defects as $\mathcal{W}\equiv\{w_k|k=1,\ldots,N\}$. 
    We also define the set of topological defects $\mathcal{Z}\equiv \{z_k|k=1,\ldots,N\}$, $z_k=g^{-1}(w_k)$, mapped from $\mathcal{W}$ and
    the small disc with radius $\epsilon$ which encloses only $w_k$ as $D_k(\epsilon)\equiv\{w_k+re^{\rmi\theta}|\ 0\leq r\leq \epsilon,0\leq \theta\leq 2\pi\}$. 
    The sum of topological charges satisfies~(\ref{eq:cond_sumq}). 
    The total energy $F$ of the region $\Omega\backslash D(\epsilon)$, $D(\epsilon)\equiv \cup_{k=1}^N D_k(\epsilon)$, is written as follows: 
    \begin{align}
        F =-2\pi \sum_{k=1}^N q_k^2 \log \epsilon +  2\pi F_d + F_g + \mathcal{O}(\epsilon\log\epsilon),\label{eq:robin_arbitrary}
    \end{align}
    where 
    \begin{align}
        &F_d \equiv \sum_{k=1}^N\left(q_k^2 - 2q_k\right)\log |g'(z_k)| + \sum_{k=1}^N q_k^2 {\rm Im}[\mathcal{R}(z_k)] + \sum_{k=1}^N\sum_{l\neq k}^N q_k q_l {\rm Im}[\mathcal{G}(z_k,z_l)],\label{eq:defect_conf}\\
        &F_g \equiv \int_{\mathbb{D}}\left|\frac{g''(z)}{g'(z)}\right|^2 \d S + 4\pi \log |g'(0)|.
    \end{align}
    The energy $F_d$ depends on $z_k$ and the map $g(z)$, whereas $F_g$ is independent of $z_k$. The energy $F_g$ can be seen as the total energy of the map $g(z)$ in $\mathbb{D}$. 
\end{theorem}

\begin{figure}[t]
        \centering
        \includegraphics[bb=400 0 700 450,scale=0.35]{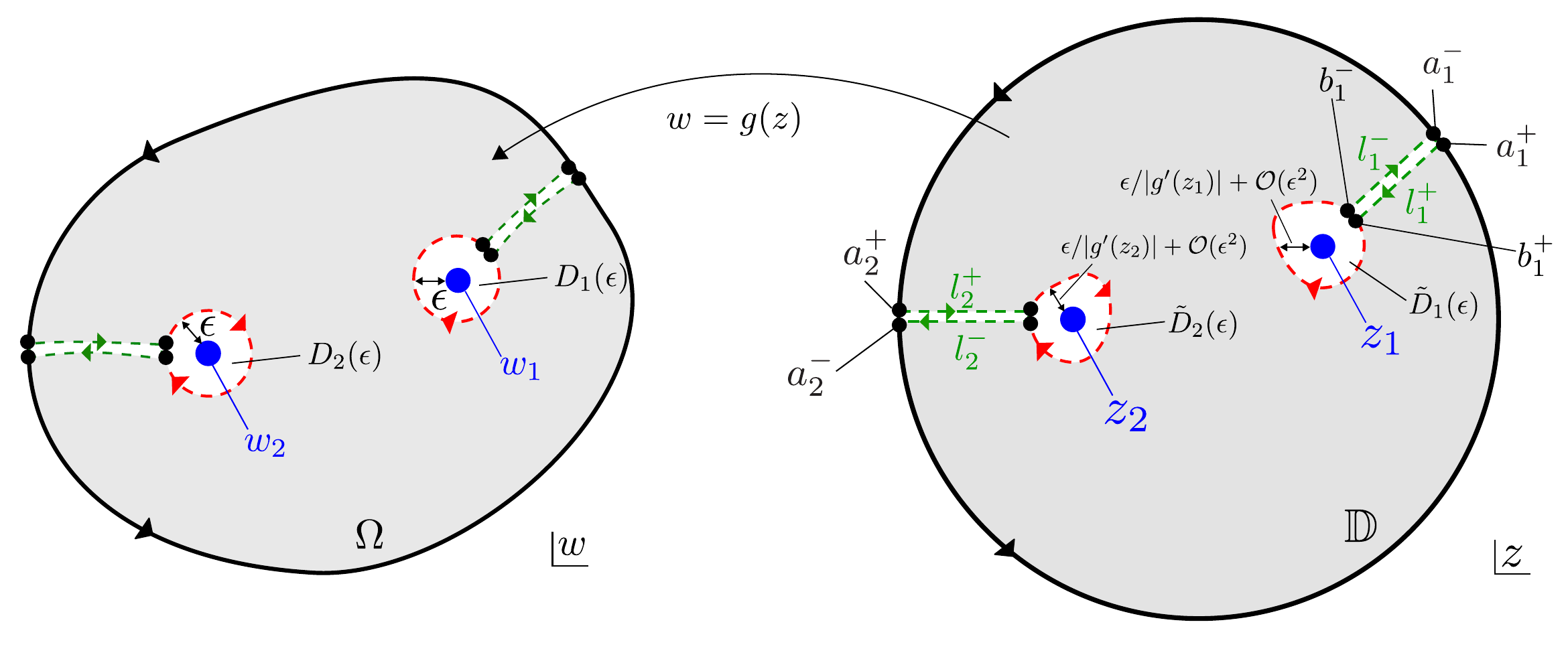}
        \caption{Line integral along $\partial (\Omega\backslash D(\epsilon))$ for case $N=2$. Branch cuts of logarithmic singularities $z_k$ are chosen as $l_k$ and argument of $\log(z-z_k)$ jumps on $l_k$ by $2\pi$. 
        The path of integrals in the $w$-plane is transformed into the line integral in the $z$-plane under a conformal map $w=g(z)$. }
        \label{fig:lineint_arbitray}
\end{figure}

\begin{proof}
    Since the energy defined by equation~(\ref{eq:Frank}) is conformally invariant, 
    the Frank free energy in the domain $\Omega$ is expressed by line integrals:
\begin{align}
    F &= \frac{1}{2\rmi}\oint_{\partial(\Omega\backslash D(\epsilon))} \frac{\partial f_\Omega}{\partial w}(g^{-1}(w)|\mathcal{Z})\conj{f_\Omega(g^{-1}(w)|\mathcal{Z})}\d w = \frac{1}{2\rmi}\oint_{\partial(\mathbb{D}\backslash\tilde{D}(\epsilon)) } \frac{\partial f_\Omega}{\partial z}(z|\mathcal{Z}) \conj{f_\Omega(z|\mathcal{Z})}\d z\nonumber\\
    &=\frac{1}{2\rmi}\left[\sum_{k=1}^N \int_{a_k^-}^{a_{k+1}^+} \frac{\partial f_\Omega}{\partial z}\conj{f_\Omega} \d z + \sum_{k=1}^N \left(\int_{l_k^+} \frac{\partial f_\Omega}{\partial z}\conj{f_\Omega} \d z -  \int_{l_k^-} \frac{\partial f_\Omega}{\partial z}\conj{f_\Omega} \d z - \oint_{\partial \tilde{D}_k(\epsilon) } \frac{\partial f_\Omega}{\partial z}\conj{f_\Omega} \d z \right)\right], \label{eq:line_intC0_all_map}
\end{align}
where the points $a_k^{\pm}$, $b_k^{\pm}$, and the paths $l_k^{\pm}$ are chosen in the same manner as that in the unit disc case 
and $\tilde{D}_k(\epsilon)\equiv g^{-1}(D_k(\epsilon))$. 
Figure~\ref{fig:lineint_arbitray} shows the paths of the line integrals. 
The point on $\partial D_k(\epsilon)$ is parameterized as $w=w_k+\epsilon e^{\rmi\theta}$, and $\theta\in[0,2\pi]$, which is transformed into 
\begin{align}
    g^{-1}(w_k + \epsilon e^{\rmi\theta}) = z_k +\frac{1}{g'(z_k)}\epsilon e^{\rmi\theta} + \mathcal{O}(\epsilon^2)
\end{align}
under the inverse of the conformal map of $g(z)$. 
This implies that the disc at $w=w_k=g(z_k)$ with radius $\epsilon$ is mapped to the disc at $z=z_k$ with radius $\tilde{\epsilon}\equiv\epsilon/|g'(z_k)|$ up to $\mathcal{O}(\epsilon^2)$.  

Now we use the same technique as that used for the unit disc. 
Before proving this, it is convenient to define $h(z) \equiv -\rmi \log g'(z)$ and a function that excludes only the singularity at $z=z_k$ as
\begin{align}
    \tilde{r}_k(z) \equiv f_\Omega(z|\mathcal{Z}) + \rmi \log (z-z_k) =  r_k(z) + h(z).
\end{align}

\textbf{(i) Line integral along $\partial \tilde{D}_k(\epsilon)$}

The integral along $\partial \tilde{D}_k(\epsilon)$ follows the calculations for the unit disc except for two simple modifications, namely, 
the inner radius $\epsilon$ on the $w$-plane is $\tilde{\epsilon}$ on the $z$-plane and the function $r_k(z)$ analytic at $z=z_k$ is $\tilde{r}_k(z)$ in this case. 
Following equation~(\ref{eq:int_Dk}), the line integral around $\tilde{D}_k(\epsilon)$ is
\begin{align}
&-\frac{1}{2\rmi}\oint_{\partial \tilde{D}_k(\epsilon)} \frac{\partial f_\Omega}{\partial z}\conj{f_\Omega}dz  = -\pi q_k^2 \log \tilde{\epsilon} + \rmi\pi  q_k \conj{\tilde{r}_k(z_k)} + \rmi \pi^2 q_k^2 + \rmi  \pi q_k^2 \beta_k^-  + \mathcal{O}(\epsilon\log\epsilon)\nonumber\\ 
&=-\pi q_k^2 \log \epsilon + \pi q_k^2 \log|g'(z_k)| + \rmi \pi q_k \conj{r_k(z_k)}- \pi q_k \log \conj{g'(z_k) } + \rmi \pi^2 q_k^2 + \rmi\pi q_k^2 \beta_k^- + \mathcal{O}(\epsilon\log\epsilon).\nonumber 
\end{align}

\vspace{3pt}

\textbf{(ii) Integrals along branch cuts $l_k^+$ and $l_k^-$}

This integral is calculated using the same technique as that used for the unit disc: 
\begin{align}
    &\frac{1}{2\rmi}\int_{l_k^+}\frac{\partial f_\Omega}{\partial z} \conj{f_\Omega} \d z - \frac{1}{2\rmi}\int_{l_k^-}\frac{\partial f_\Omega}{\partial z} \conj{f_\Omega} \d z \nonumber \\
    &=-\pi q_k^2 \log \epsilon + \pi q_k^2 \log |g'(z_k)| - \rmi\pi q_k^2 \beta_k^- - \rmi \pi  q_k r_k(z_k)-\pi q_k \log g'(z_k) + \rmi\pi q_k f_\Omega(a_k^- |\mathcal{Z}) + \mathcal{O}(\epsilon). \nonumber
\end{align}

\vspace{3pt}

\textbf{(iii) Line integral along the unit circle}

The line integral along the unit circle is given by 
\begin{align}
\frac{1}{2\rmi}\sum_{k=1}^N \int_{a_k^-}^{a_{k+1}^+} \frac{\partial f_\Omega}{\partial z}\conj{f_\Omega} \d z &= \frac{1}{2\rmi}\sum_{k=1}^N \int_{a_k^-}^{a_{k+1}^+}\left(\frac{\partial f_\mathbb{D}}{\partial z} + \frac{\partial h}{\partial z} \right)\conj{(f_\mathbb{D}+h)}\d z\nonumber \\
&=\frac{1}{2\rmi}\sum_{k=1}^N \int_{a_k^-}^{a_{k+1}^+}\left(\frac{\partial f_\mathbb{D}}{\partial z}\conj{f_\mathbb{D}} + \frac{\partial f_\mathbb{D}}{\partial z}\conj{h} + \frac{\partial h}{\partial z}\conj{f_\mathbb{D}} + \frac{\partial h}{\partial z}\conj{h} \right)\d z.\label{eq:all_arb}
\end{align}
Using the same argument as that for the unit disc, the first term in (\ref{eq:all_arb}) is identical to~(\ref{eq:around_C0final}); that is, 
\begin{align}
    \frac{1}{2\rmi}\sum_{k=1}^N \int_{a_k^-}^{a_{k+1}^+}\frac{\partial f_\mathbb{D}}{\partial z}\conj{f_\mathbb{D}}\d z = -\rmi \pi \sum_{k=1}^N (q_k f_\mathbb{D}(a_k^{-}|\mathcal{Z}) + \pi q_k^2).
\end{align}
The last term in (\ref{eq:all_arb}) is 
\begin{align}
    \frac{1}{2\rmi}\sum_{k=1}^N \int_{a_k^-}^{a_{k+1}^+}\frac{\partial h}{\partial z}\conj{h}\d z = \int_{\mathbb{D}}\left|\frac{g''}{g'} \right|^2\d x \d y,
\end{align}
where the complex Green's theorem and the fact that $h(z)$ is analytic in $\mathbb{D}$ are used. 
The second term in~(\ref{eq:all_arb}) is
\begin{align}
    \frac{1}{2\rmi}\sum_{k=1}^N \int_{a_k^-}^{a_{k+1}^+}\frac{\partial f_\mathbb{D}}{\partial z} \conj{h} \d z = -\frac{1}{2}\oint_{\partial \mathbb{D}} \sum_{k=1}^N q_k\left(\frac{1}{z-z_k} - \frac{\conj{z_k}}{1-\conj{z_k}z} \right)\conj{h}\d z =\rmi \pi \sum_{k=1}^N q_k \conj{h(z_k)} - 2\rmi \pi \conj{h(0)},    
\end{align}
where we used $\conj{z}=1/z$, $\d\conj{z}=-1/z^2 \d z$ on $\partial \mathbb{D}$, and the following residue theorem:
\begin{align}
    &\frac{1}{2}\sum_{k=1}^N \int_{a_k^-}^{a_{k+1}^+} \frac{\conj{h(z)}}{z-z_k}\d z = \frac{1}{2}\oint_{\partial \mathbb{D}} \frac{\conj{h(z)}}{z-z_k} \d z = \conj{\frac{1}{2}\oint_{\partial \mathbb{D}} \frac{h(z)}{1/z-\conj{z_k}}\left(-\frac{1}{z^2}\right) \d z}  = \rmi\pi \conj{h(0)},\\
    &-\frac{\conj{z_k}}{2}\sum_{k=1}^N \int_{a_k^-}^{a_{k+1}^+} \frac{\conj{h(z)}}{1-\conj{z_k}z}\d z = -\frac{\conj{z_k}}{2} \conj{\oint_{\partial \mathbb{D}} \frac{h(z)}{1-z_k/z}\left(-\frac{1}{z^2}\right) \d z} = -\rmi\pi\conj{h(z_k)} + \rmi\pi \conj{h(0)}.
\end{align}

The third term is calculated based on the property that $\conj{f_\mathbb{D}} = 2\theta + 2\chi(\theta) - \pi - f_\mathbb{D}$ on the unit circle:
\begin{align}
    \frac{1}{2\rmi}\sum_{k=1}^N \int_{a_k^-}^{a_{k+1}^+}\frac{\partial h}{\partial z}\conj{f_\mathbb{D}}\d z= \frac{1}{2\rmi}\sum_{k=1}^N \int_{a_k^-}^{a_{k+1}^+}\frac{\partial h}{\partial z}(2\theta + 2\chi(\theta) - \pi - f_\mathbb{D})\d z.\label{eq:hf}
\end{align}
The first two terms in (\ref{eq:hf}) are explicitly calculated as follows:
\begin{align}
    \frac{1}{\rmi}\sum_{k=1}^N \int_{a_k^-}^{a_{k+1}^+}\frac{\partial h}{\partial z} (\theta+\chi(\theta)) \d z = -2\rmi\pi \sum_{k=1}^N q_k h(a_k^{-}) + \rmi\int_{0}^{2\pi} h(e^{\rmi\theta}) \d \theta= -2\rmi\pi \sum_{k=1}^N q_k h(a_k^{-}) + 2\rmi\pi  h(0).\nonumber 
\end{align}
The third term in (\ref{eq:hf}) is zero because $h(z)$ is analytic in $\mathbb{D}$. The fourth term in (\ref{eq:hf}) is 
\begin{align}
    -\frac{1}{2\rmi}\sum_{k=1}^N \int_{a_k^-}^{a_{k+1}^+} \frac{\partial h}{\partial z} f_\mathbb{D} \d z &=  -\frac{1}{2\rmi}\sum_{k=1}^N[h(z) f_\mathbb{D}]_{a_k^-}^{a_{k+1}^+} + \frac{1}{2\rmi}\sum_{k=1}^N\int_{a_k^-}^{a_{k+1}^+} \frac{\partial f_\mathbb{D}}{\partial z} h \d z \nonumber \\
    &=\rmi \pi\sum_{k=1}^N  q_k h (a_k^{-}) - \rmi \pi\sum_{k=1}^N q_k h(z_k),
\end{align}
where we used the residue theorem. Hence, the line integral along the unit circle is given by
\begin{align}
    \frac{1}{2\rmi}\sum_{k=1}^N \int_{a_k^-}^{a_{k+1}^+} \frac{\partial f_\Omega}{\partial z}\conj{f_\Omega} \d z &= -\rmi\pi\sum_{k=1}^N q_k h(a_k^{-}) - 2\rmi \pi \sum_{k=1}^N q_k {\rm Im}[h(z_k)] + 4\pi \log |g'(0)|\nonumber\\
    &-\rmi \pi\sum_{k=1}^N q_k f_\mathbb{D}(a_k^{-}|\mathcal{Z})- \rmi \pi^2 \sum_{k=1}^N q_k^2 + \int_{\mathbb{D}}\left|\frac{g''}{g'} \right|^2 \d x \d y.
\end{align}
Combining (i), (ii), and (iii), the energy for $f_\Omega(z|\mathcal{Z})$ given by~(\ref{eq:robin_arbitrary}) is derived.

\end{proof}

\section{Topological defects localization on quadrature domains} \label{sec:4}

{\color{black}
Section~\ref{sec:3} presented the explicit formula for the Frank free energy of arbitrarily shaped domains.  
The complex potential $f_\Omega$ was split into analytic and logarithmic parts and the line integrals along boundaries were evaluated up to $\epsilon\log\epsilon$. 
The formulas derived here are explicit, written as logarithmic singularities plus the contribution from conformal maps. 
This section applies the formulas~(\ref{eq:robin_arbitrary}) to validate a conjecture given by Ienaga {\em et al.}~\cite{Ienaga2023-qm}, namely, 
that the positions of defects in both doublets and triplets linearly correlate with the distances between the centres of the overlapping discs. 

The simplest extension of localizing topological defects with boundaries, as done by Duclos~{\em et al.}~\cite{Duclos2016-gn},
is to estimate defect positions in regions with multiple discs that intersect each other. }
When two or three discs with the same radius overlap at a distance $\Delta$, they are called ``doublets'' and ``triplets'', respectively. 
These geometries exhibit ferromagnetic and antiferromagnetic order in bacterial vortex lattices, 
depending on the distance between the centres of the discs relative to the radius of the discs~\cite{wioland2016ferromagnetic}. 
Ienaga {\em et al}. created regions with two or three circular areas and 
conducted cell alignment experiments within these regions to investigate the number and positioning of defects~\cite{Ienaga2023-qm}. 
They hypothesized a proportional relationship between the distance $\Delta$ between the centres of two overlapping discs and 
the distance $D_{++}$ between two $+1/2$ defects within those doublets. They conjectured the following equation for doublets through fitting:
\begin{align}
    D_{++} = 2\left(D_{++,0} + \frac{\Delta}{2}\right),\quad D_{++,0} = 0.58.
\end{align}
For triplets, it was experimentally confirmed that 
when $\Delta$ is large, a topological defect with a charge of $-1/2$ exists at the centre, and three topological defects with a charge of $+1/2$ are located on each disc. 
They also hypothesized that the distance $D_{+-}$ between the $-1/2$ defect and the $+1/2$ defect, 
and the distance between disc centres $\Delta$, have a proportional relationship with a slope of $1/\sqrt{3}$.

In this paper, 
we analytically verify these hypotheses using the proposed formula and the mapping of quadrature domains~\cite{gustafsson2005quadrature}. 
The conformal mapping for a region with $M$-overlapping unit discs can be approximated by the following rational function:
\begin{align}
    w = g(z) = \frac{a^{2M}-1}{a^M - a^{M-1}+1}\cdot \frac{z}{a^M - z^M}, \label{eq:quadrature_gz}
\end{align}
where the real parameter $a$ should be greater than $1$ and $M$ is a positive integer. 
This region is known as the quadrature domains. 
Such domains have been extensively studied in fields such as Hele-Shaw flows~\cite{crowdy2020solving_book,gustafsson2005quadrature}.

\begin{figure}[t]
    \centering
    \includegraphics[bb=250 0 1300 370,scale=0.32]{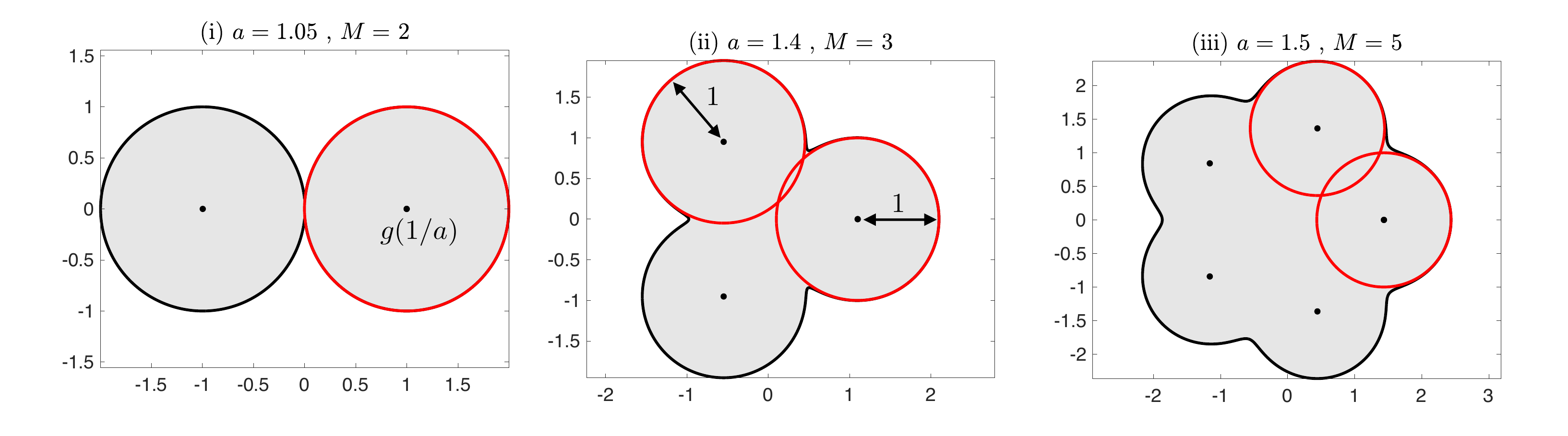}
    \caption{Examples of quadrature domains with various values of $M$ and $a$. 
    Black dots correspond to $w=g(\omega_M^m/a)$, $\omega_M\equiv \exp(2\pi\rmi/M)$, $m=0,1,\ldots,M-1$, which are considered as the centres of the discs. 
    Red circles are unit discs with centres at the black dots. 
    (i) $a=1.05$, $M=2$, (ii) $a=1.4$, $M=3$, and (iii) $a=1.5$, $M=5$. }
    \label{fig:quad}
\end{figure}
\begin{figure}[t]
    \centering
    \includegraphics[bb=-120 0 1500 480,scale=0.44]{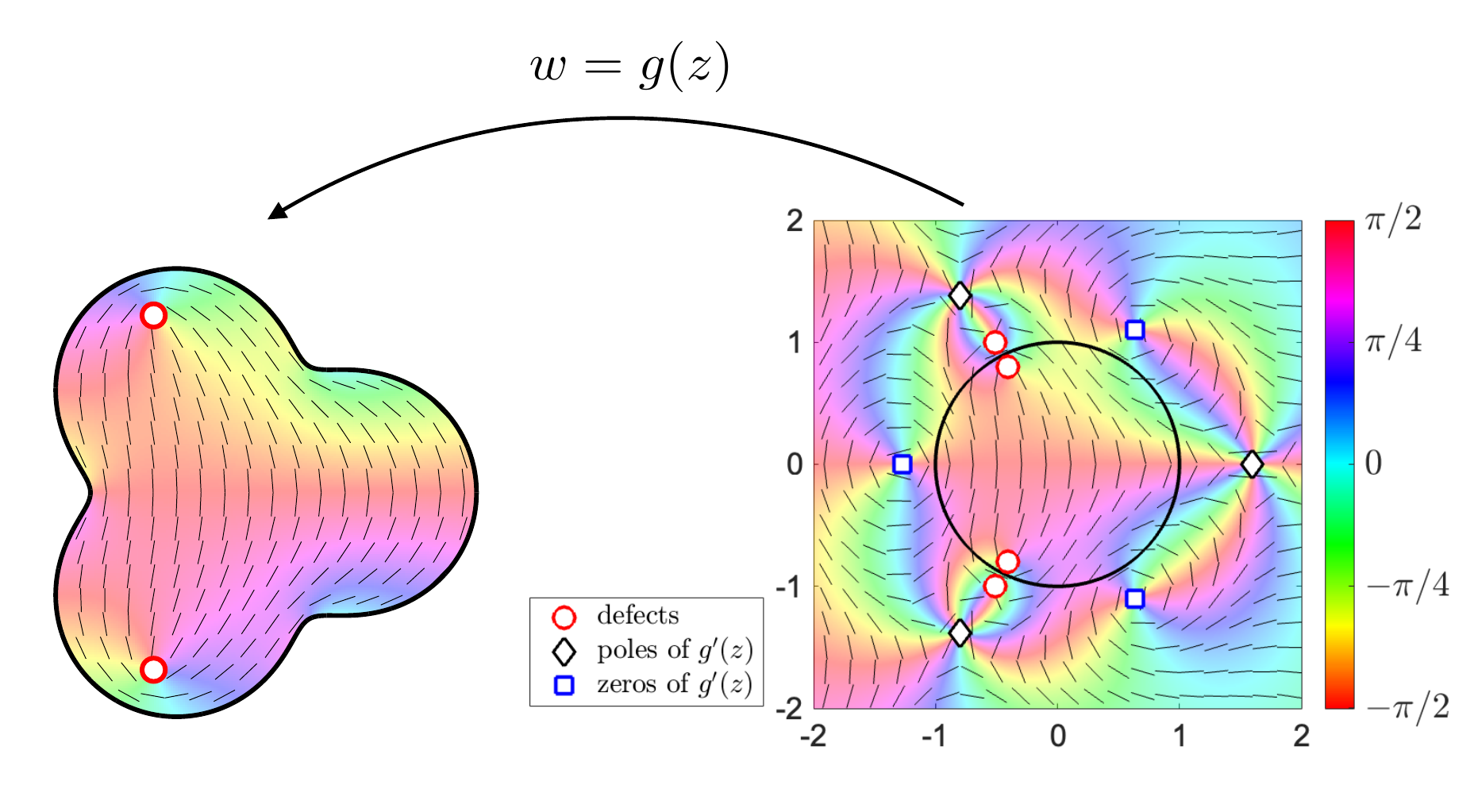}
    \caption{Example of quadrature domain with $M=3$ and two stationary topological defects. 
    Two topological defects with a charge of $+1/2$ exist inside the unit disc. 
    Because of the potential $f_\Omega$, topological defects with a charge of $+1/2$ are also located at $1/\conj{z_n}$, $n=1,2$. 
    For $M=3$, $g'(z)$ has three simple zeros and three double poles outside the unit disc, which indicates that the correction term $h(z)=-\rmi\log g'(z)$ produces topological defects with a charge of $+1$ and $-2$. 
    The problem of finding the stationary positions of defects is equivalent to finding an equilibrium of ten topological defects in the infinite plane.}
    \label{fig:tsuchi}
\end{figure}

Quadrature domains are applied to calculate topological defects for the first time in the field of active nematics. 
The advantage of using the rational function~(\ref{eq:quadrature_gz}) is 
that it enables the boundary of doublets or triplets to be represented by changing a single parameter $a$. 
Figure~\ref{fig:quad} shows the results of mappings for various values of $M$ and $a$. 
Through this mapping, the origin of the unit disc in the $z$-plane is mapped to the origin of the $w$-plane, corresponding to the centre of the region in the $w$-plane. 
The first and second derivatives of this function are given as follows:
\begin{align}
    &g'(z) = \frac{a^{2M}-1}{a^M - a^{M-1} + 1 }\cdot\frac{(a^M + (M-1) z^M)}{(a^M - z^M)^2},\\
    &g''(z) = \frac{a^{2M}-1}{a^M - a^{M-1} + 1 }\cdot\frac{Mz^{M-1}[(M-1)z^M + (M+1)a^M]}{(a^M-z^M)^3}.
\end{align}
The mapping $g(z)$ may exhibit self-intersections for certain value of $a$. 
The condition for no self-intersections is that the turning number of $g(z)$ around the unit disc is $1$, that is,  
\begin{align}
    \frac{1}{2\pi}\int_{0}^{2\pi} \frac{\d}{\d\theta} {\rm Im}\left[\log g'(e^{\rmi\theta}) \right] \d \theta = 1,
\end{align}
which results in no zeros of $g'(z)$ being in the unit disc, i.e., $a>(M-1)^{1/M}$. 

The centres of the discs are well approximated by $g(\omega^{m}_M/a)$, $m=0,1,\ldots,M-1$, $\omega_M=\exp(2\pi\rmi/M)$. 
The coefficients of the mapping in (\ref{eq:quadrature_gz}) are chosen such that the radius $R=g(1)-g(1/a)$ 
equals $1$. 

The advantage of the map is that $g(z)$ is rational, allowing the Frank free energy in the presence of topological defects to be expressed as logarithm of certain rational functions. 
This enables the calculation of stationary defect locations and stability using standard calculus. 
We note that 
\begin{align}
    -\rmi \log g'(z) = -\rmi \log (M-1) -\rmi \sum_{m=1}^M \log(z+a\omega_M^m/(M-1)^{1/M}) + 2\rmi \sum_{m=1}^M \log(z-a\omega_M^m),
\end{align}
where $\omega_M \equiv \exp(2\pi\rmi/M)$. 
This indicates that the problem of finding two stationary defects considered here is 
to find the stability of four topological defects with a charge of $+1/2$, $M$ topological defects with a charge of $+1$ located at $z=-a\omega_M^m/(M-1)^{1/M}$, $m=0,\ldots,M-1$, and  topological defects with a charge of $-2$ located at $z=a \omega^{m}_{M}$, $m=0,\ldots,M-1$ in the infinite plane. 
The configuration for $M=3$ is illustrated in Figure~\ref{fig:tsuchi}.

\subsection{$M=2$: Doublet domain}
We first investigate the defect positions and 
their alignment angles in doublet regions. 
The condition for no self-intersections is $a>1$ and 
\begin{align}
    g(z) = \frac{a^4-1}{a^2 - a + 1}\cdot \frac{z}{a^2 -z^2}.
\end{align}
Since the centres of the discs of doublet regions are represented as $g(\pm 1/a)$, the distance between the centres of the two discs is given by 
\begin{align}
    \Delta = 2g(1/a) = \frac{2a}{a^2 - a + 1}.
\end{align}
Using the symmetry in the region where the two discs intersect, 
we assume that there exist topological defects with charges $q_1=q_2=+1/2$ at two symmetric points $\pm g(s)$ along the $u$-axis of the $w=u+\rmi v$-plane, where $0<s<1$. 
In other words, the positions of topological defects in the predomain are $z=\pm s$. 
Using equation~(\ref{eq:robin_arbitrary}), the free energy that depends solely on the positions of the defects is explicitly given by 
\begin{align}
    F_d &=-\frac{1}{2}\log[2s(1-s^4)] - \frac{3}{4}\log|g'(s)| - \frac{3}{4}\log|g'(-s)|.
\end{align}
To find the extrema of the free energy, consider the derivative with respect to $s$:
\begin{align}
    \frac{\partial F_d}{\partial s} = -\frac{3g''(s)}{4g'(s)} + \frac{3g''(-s)}{4g'(-s)} + \frac{5s^4-1}{2s(1-s^4)} = \frac{3s(s^2 + 3a^2)}{s^4 - a^4} + \frac{5s^4-1}{2s(1-s^4)}=0.
\end{align}
This gives an $8$th-degree polynomial equation with only even powers of $s$:
\begin{align}
    p_2(t) = t^4 + 18a^2 t^3 + 5(a^4-1)t^2 - 18a^2 t - a^4 = 0,\quad t = s^2. \label{eq:polynomial}
\end{align}
Using the fact that $p_2(0)=-a^4<0$ and $p_2(1)=4(a^4-1)>0$, $p_2(t)$ has at least one real solution in the interval $0<t<1$, and 
\begin{align}
    &p_2'(t)=4 t^3 + 54 a^2 t^2 + 10(a^4-1)t - 18a^2,\\
    &p_2''(t)= 12t^2 + 108 a^2 t + 10(a^4-1)>0,\quad 0<t<1,
\end{align}
which indicates that $p_2(t)$ has only one real solution in the interval $0<t<1$. 
We denote this solution as $t_0$ and define $s_0^2= t_0$, $s_0>0$. 
We solve the polynomial equation~(\ref{eq:polynomial}) numerically between $0<s<1$. 
We refer to this method as the polynomial method.

\begin{figure}[t]
    \centering
    \includegraphics[bb=50 0 1300 410   ,scale=0.35]{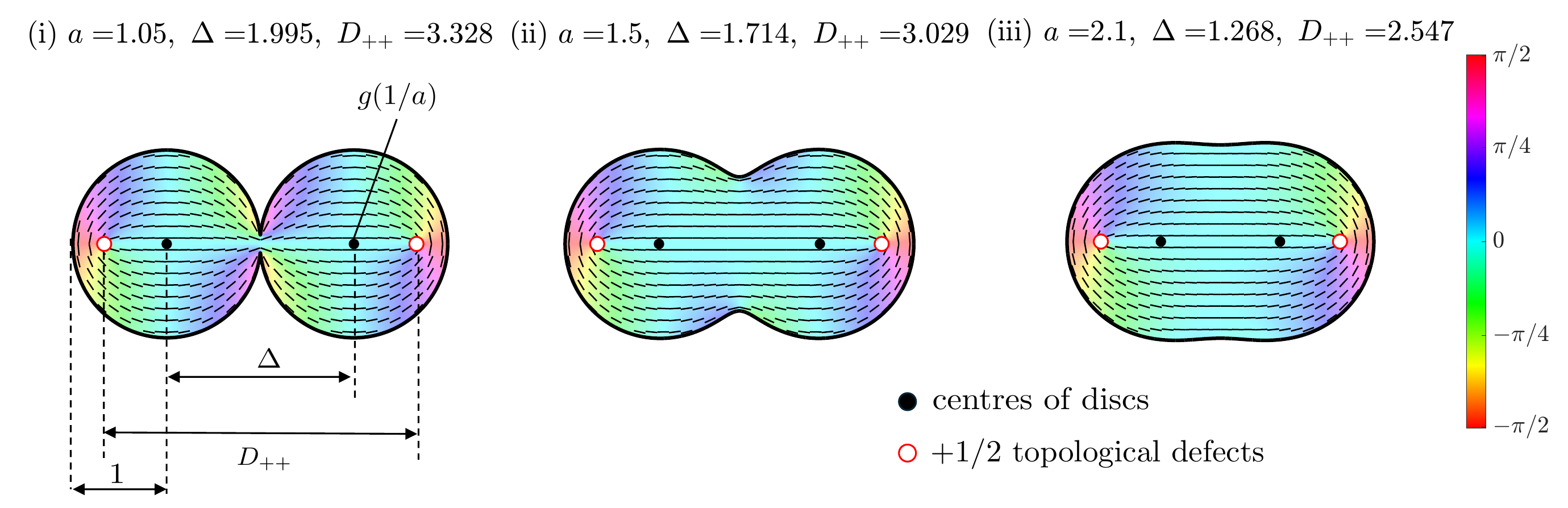}
    \caption{Colour map of $\phi$ created by $+1/2$ defect pairs (red circles) in quadrature domain with $M=2$.
    The defect locations are determined through energy minimization under the assumption that two $+1/2$ defects symmetrically located at $z=\pm s$ on the real axis of the predomain. 
    Black dots correspond to $w=g(1/a)$, which are assumed to be the centres of the discs of doublets based on numerical experiments. }
    \label{fig:quad2}
\end{figure}

\begin{figure}[t]
    \centerline{\includegraphics[bb=50 0 550 250,scale=0.68]{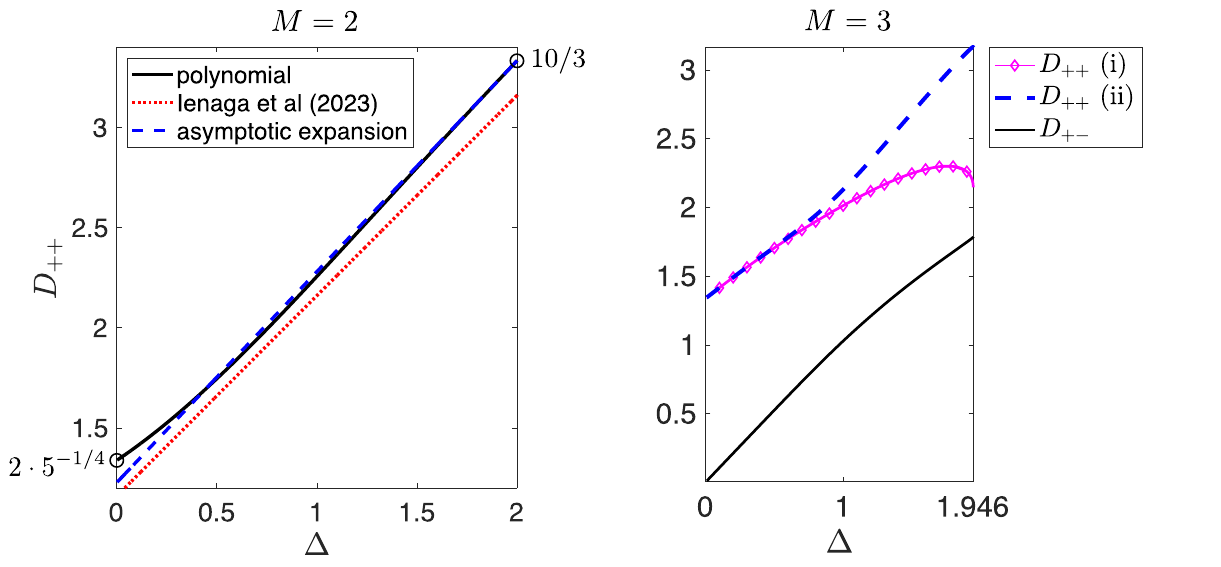}}
    \caption{(Left) Relationship between distance between two discs $\Delta$ and distance between two defects $D_{++}$. 
    Black line is the numerical results obtained using the polynomial method and the red dotted line corresponds to the conjecture given in~\cite{Ienaga2023-qm}. 
    The polynomial is well approximated by the blue dotted line given by asymptotic expansion near $\Delta = 2$. (Right) 
    Relationship between $\Delta$ and $D_{++}$, $D_{+-}$ for triplets. We checked that the slope of the black line at $\Delta = 1.9$ is $0.7232$, which is close to the value of $1/\sqrt{3}$ conjectured in~\cite{Ienaga2023-qm}. 
    Our numerical results are within the error bars in~\cite{Ienaga2023-qm}. }
    \label{fig:linear}
\end{figure}

Figure~\ref{fig:quad2} shows the locations of topological defects and colour maps of $\phi$ in $M=2$ with (i) $a=1.05$, (ii) $=1.5$, and $a=2.1$. 
The defect locations are determined by using the polynomial method. 
Figure~\ref{fig:linear} shows numerical results for the distance between two defects $D_{++}$ 
for various values of $\Delta$. 
The solid black line represents the results obtained using the polynomial method. 
The results conjectured by~\cite{Ienaga2023-qm} (red dashed line) are in good agreement with our numerical results. 
However, it can be observed that the formula expected by Ienaga {\em et al.} is slightly shifted towards smaller values.

\subsection{Asymptotic expansions of $D_{++}$ and $\Delta$ for $M=2$}
We demonstrate that the distance between two defects is proportional to the distance between the centres of the two circles in the limit 
$a\rightarrow 1$ (i.e., when the two discs are almost touching each other). 
For $a=1+k$ with a small parameter $k$, the distance $\Delta$ is expanded with respect to $k$ as follows:
\begin{align}
    \Delta = \frac{2a}{a^2 - a+1} = 2(1-k^2 + k^3 ) + \mathcal{O}(k^4).
\end{align}
The expansion of the root of the polynomial~(\ref{eq:polynomial}) is given by 
\begin{align}
    s_0 = 1 - \frac{1}{5}k + \frac{3}{25}k^2- \frac{99}{625}k^3+ \frac{32}{125}k^4+ \mathcal{O}(k^5),
\end{align}
which gives 
\begin{align}
    D_{++} &= 2g(s_0) = \frac{10}{3} - \frac{158}{75}k^2 + \frac{2576}{1125}k^3 + \mathcal{O}(k^4)\nonumber \\
    &=\frac{92}{75} + \frac{79}{75}\Delta + \frac{206}{1125}k^3 + \mathcal{O}(k^4).\label{eq:proportional}
\end{align}
It is important to note that $92/75 = 1.227 \sim 2D_{++,0} = 1.16$, which well agrees with the formula conjectured by Ienaga~{\em et al}. 
It is also observed that $D_{++}$ and $\Delta$ are proportional for sufficiently small values of parameter $k$ and that 
the approximation accuracy is good because the coefficient of $k^3$ is small. 
The proportional equation, which includes only the first two terms in~(\ref{eq:proportional}) here,
is referred to as asymptotic expansion. 

The blue dotted line in Figure~\ref{fig:linear} shows the result of asymptotic expansion. 
This line well matches the numerical results for wide ranges of $\Delta$. 

\subsection{$M=3$: Triplet domain}
Ienaga {\em et al}. also investigated the positions of defects and the distances between the centres of three overlapping discs (i.e., triplets)~\cite{Ienaga2023-qm}. 
They experimentally demonstrated that, in this case as well, 
there exists a proportionality between the positions of defects and the circular regions.

\begin{figure}[t]
    \centerline{\includegraphics[bb=-150 0 1600 1370,scale=0.3]{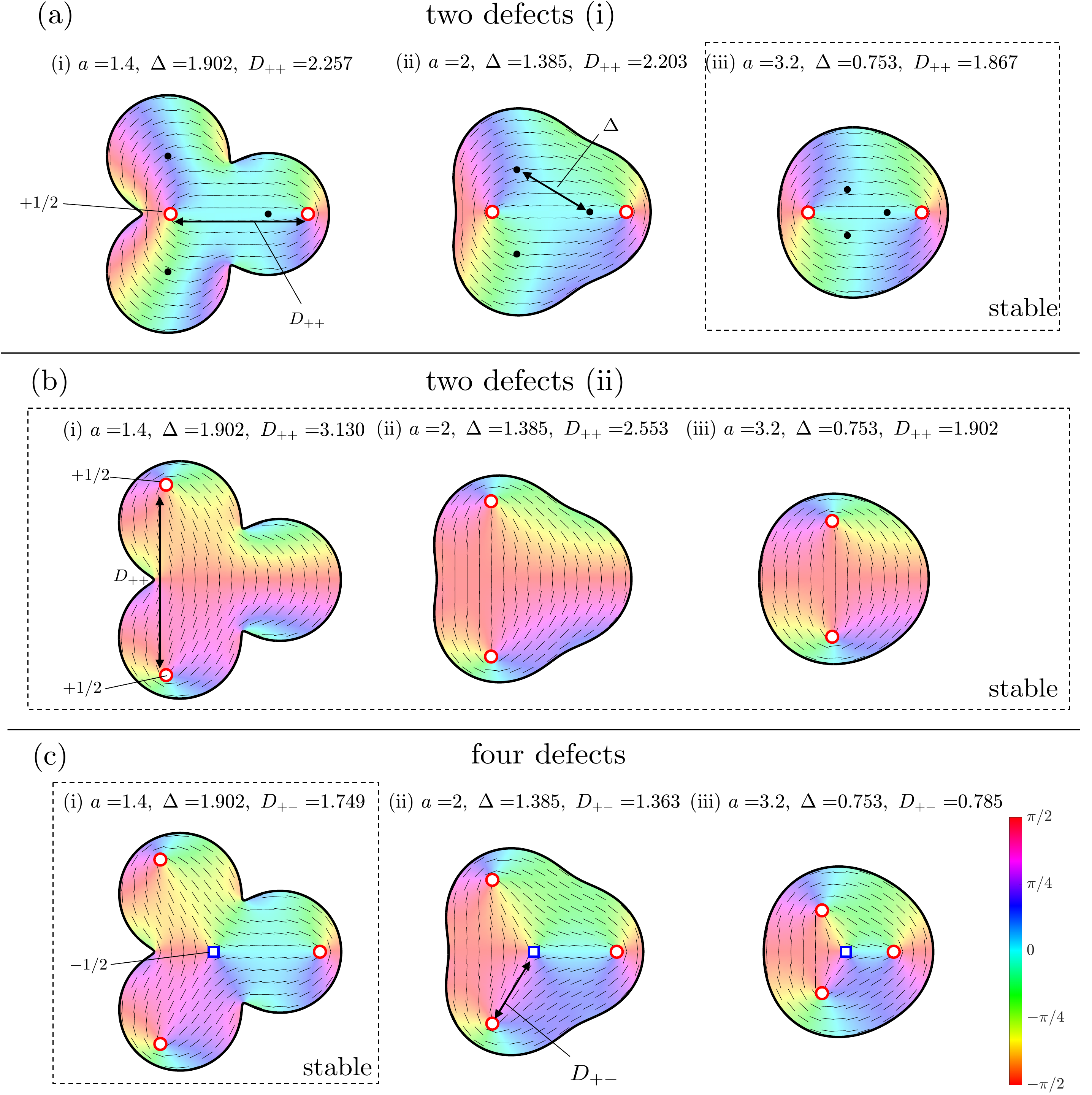}}
    \caption{Defect positions in triplets. Black dots in (a) correspond to the centres of the circles. 
    Red circles are topological defects with a charge of $+1/2$ and blue squares are defects with a charge of $-1/2$. 
    (a) Case where two defects exist on the real axis, 
    (b) case where defects are located at positions symmetric about the $x$-axis, and 
    (c) case where defects with a charge of $+1/2$ are located at each disc and the one defect with a charge of $-1/2$ is located at the origin.}
    \label{fig:three}
\end{figure}

For $M=3$, 
\begin{align}
    g(z) = \frac{a^6 - 1}{a^3 - a^2 + 1}\cdot \frac{z}{a^3 - z^3},\quad g'(z) = \frac{a^6-1}{a^3 - a^2 + 1}\cdot \frac{a^3+2z^3}{(a^3-z^3)^2},
\end{align}
where $a>2^{1/3}$ due to the condition for no self-intersections.  
Figure~\ref{fig:three} plots different shapes created by the mapping for various 
$a$. 
Ienaga {\it et al}. examined the intersections of discs and the number of defects. 
They observed that when the distances between the discs are sufficiently large, a defect with a charge of $-1/2$ forms at the center, and $+1/2$ defects form at positions 
$w = \hat{r}\exp(2\pi m \rmi/3)$, $\hat{r}\in \mathbb{R}$, $m=0,1,2$. 
Furthermore, when the three discs intersect, they experimentally found that the three $+1/2$ defects no longer exist; 
instead, one defect annihilates with the central $-1/2$ defect, 
resulting in two $+1/2$ defects forming along the $x$-axis or at positions symmetric to the $x$-axis. 

The positions of these defects are algebraically determined by assuming symmetry for the domain. 
To investigate whether the resulting defects are stable, the Hessian of the energy at these positions is examined and their stability is evaluated based on whether the Hessian is positive definite. For the calculation of the Hessian, the following quantities are precomputed:
\begin{align}
    \frac{g''}{g'} = \frac{6z^2}{a^3 + 2z^3} + \frac{6z^2}{a^3 - z^3},\quad \left(\frac{g''}{g'}\right)' = \frac{12z(a^3-z^3)}{(a^3 + 2z^3)^2} + \frac{6z(2a^3+z^3)}{(a^3 - z^3)^2}.
\end{align}
The distance between discs is given by 
\begin{align}
    \Delta = \sqrt{3}g(1/a) = \frac{\sqrt{3} a^2}{a^3 - a^2 + 1}.
\end{align}


\subsubsection*{Two defects (i): two defects located on real axis of predomain}

Assume that there exist two defects on the centre line of the quadrature domain.  
The coordinates of these defects in the predomain are defined by $z_1=r_1, \ z_2=r_2$, $-1<r_1<r_2<1$. The energy $F_d$ is explicitly given by 
\begin{align}
    F_d&=  - \frac{1}{4}\left[\log(1-r_1^2) + \log(1-r_2^2) + 2\log(r_2-r_1) + 2\log(1-r_1r_2)\right] - \frac{3}{4}[\log g'(r_1) +\log g'(r_2)].\nonumber
\end{align}
The derivative of $F_d$ with respect to $r_1$ and $r_2$ gives 
\begin{align}
    \left\{
    \begin{aligned}
    \frac{\partial F_d}{\partial r_1} &=\frac{r_1}{2(1-r_1^2)} -\frac{1}{2(r_2-r_1)} + \frac{r_2}{2(1-r_1r_2)} - \frac{3}{4}\left[\frac{6r_1^2}{a^3+2r_1^3} - \frac{6r_1^2}{r_1^3 - a^3} \right] = 0,\\
    \frac{\partial F_d}{\partial r_2} &=\frac{r_2}{2(1-r_2^2)} -\frac{1}{2(r_1-r_2)} + \frac{r_1}{2(1-r_1r_2)} - \frac{3}{4}\left[\frac{6r_2^2}{a^3+2r_2^3} - \frac{6r_2^2}{r_2^3 - a^3} \right] = 0.
    \end{aligned}
    \right. \label{eq:two_cond_r1r2}
\end{align}
The solution~(\ref{eq:two_cond_r1r2}) for $r_1$ and $r_2$ are obtained by the simplex method implemented in the standard matlab solvers~\cite{lagarias1998convergence}. 
The Hessian of the coordinates $z_1 = r_1$ and $z_2 = r_2$ is calculated as follows:
\begin{align}
    H = \begin{pmatrix}
        J& \mathcal{O}\\ \mathcal{O} &J  
    \end{pmatrix}
        \begin{pmatrix}
        H_{11} & H_{12}\\ H_{21} & H_{22}
    \end{pmatrix}
    \begin{pmatrix}
        J& \mathcal{O}\\ \mathcal{O} &J  
    \end{pmatrix}^\top,
\end{align}
where each component is given by 
\begin{align}
    \mathcal{J} \equiv  \begin{pmatrix}
        1& 1 \\ \rmi & -\rmi
    \end{pmatrix},\quad H_{kl} \equiv \begin{pmatrix}
        \dfrac{\partial^2 F_d}{\partial z_k \partial z_l} & \dfrac{\partial^2 F_d}{\partial z_k \partial \conj{z_l}} \vspace{3pt} \\ 
       \dfrac{\partial^2 F_d}{\partial \conj{z_k} \partial z_l} & \dfrac{\partial^2 F_d}{\partial \conj{z_k} \partial \conj{z_l}}
    \end{pmatrix}.
\end{align}
When the Hessian $H$ is positive definite, the coordinates $z=r_1$ and $z=r_2$ are stable.  

\begin{figure}[t]
    \centerline{\includegraphics[bb=50 0 860 420,scale=0.48]{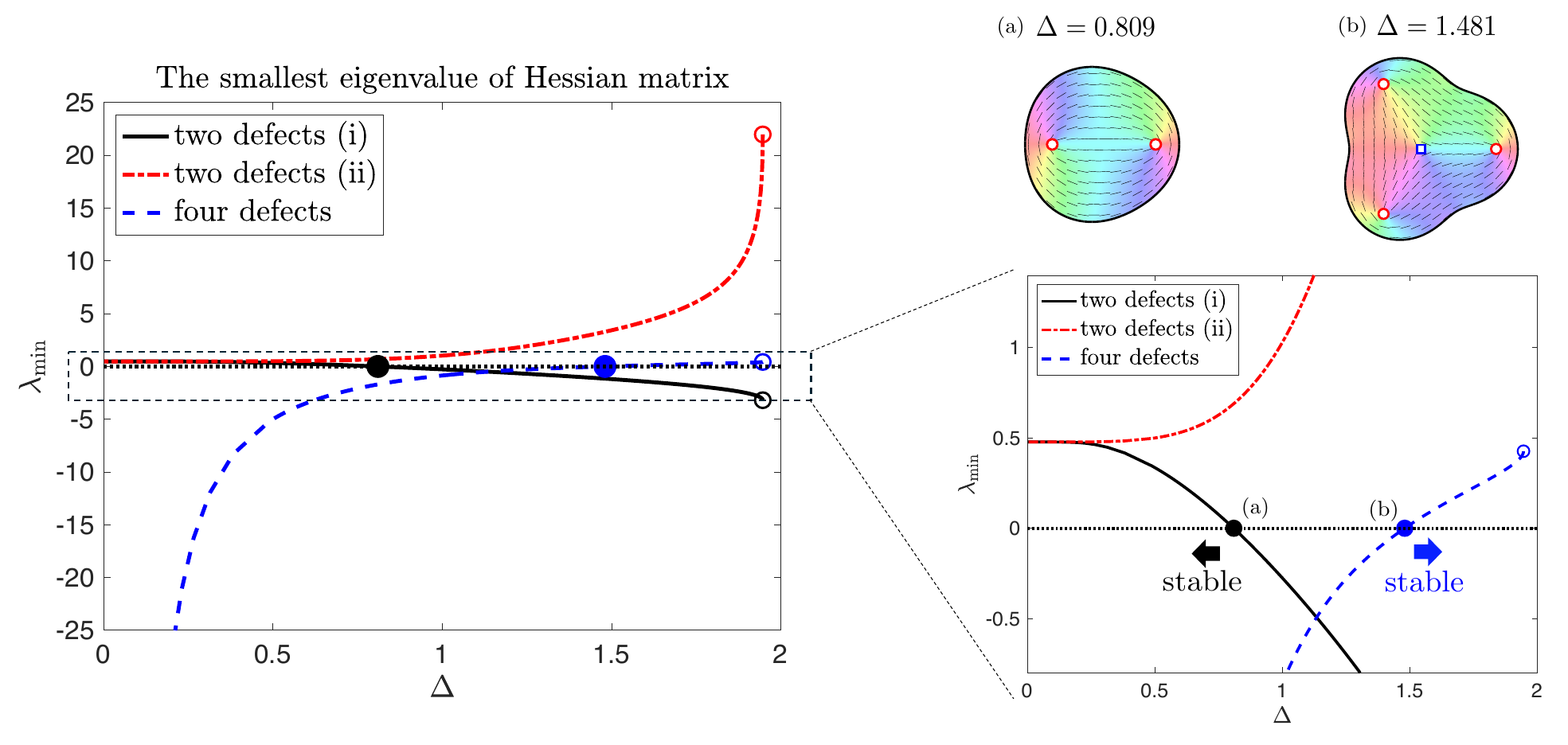}}
    \caption{Smallest eigenvalue of Hessian matrix with respect to $\Delta$. When $\Delta$ is less than $0.809$, two defects on the centre line are stable. 
    When $\Delta>1.481$, four defects are stable. 
    The formation of two defect pairs symmetrically positioned along the real axis is shown to be always stable with respect to $\Delta$.}
    \label{fig:three_two}
\end{figure}

The black solid line in Figure~\ref{fig:three_two} shows the smallest eigenvalue of the Hessian $H$. 
Our numerical simulations show that the defects become unstable when $\Delta$ becomes larger than approximately $0.809$.

\subsubsection*{Two defects (ii):  two defect pairs symmetrically positioned along real axis}
Assuming symmetry about the real axis, two defect positions in the predomain are set to $z_1 = re^{\rmi \theta}$, $z_2 = re^{-\rmi \theta}$. 
The conditions of $r$ and $\theta$ for the two defect locations are as follows: 
\begin{align}
    \left\{
    \begin{aligned}
    &\frac{\partial F_d}{\partial z_1} = -\frac{1}{4}\left(-\frac{re^{-\rmi\theta}}{1-r^2} + \frac{1}{r(e^{\rmi\theta} - e^{-\rmi\theta})} + \frac{-r e^{-\rmi \theta}}{1-r^2 e^{2\rmi\theta}} \right) - \frac{3}{8}\frac{g''(re^{\rmi\theta})}{g'(re^{\rmi\theta})}=0,\\
    &\frac{\partial F_d}{\partial z_2} = -\frac{1}{4}\left(-\frac{re^{\rmi\theta}}{1-r^2} + \frac{1}{r(e^{-\rmi\theta} - e^{\rmi\theta})} + \frac{-r e^{\rmi \theta}}{1-r^2 e^{-2\rmi\theta}} \right) - \frac{3}{8}\frac{g''(re^{-\rmi\theta})}{g'(re^{-\rmi\theta})}=0.
    \end{aligned}
    \right. \label{eq:cond_rtheta}
\end{align}
By solving (\ref{eq:cond_rtheta}) for $r$ and $\theta$ numerically subject to $0<r<1$ and $0<\theta<\pi$, 
we can find a unique solution. 
Using this solution, the Hessian is calculated in the same manner.

The red dashed line in Figure~\ref{fig:three_two} shows the smallest eigenvalue of the Hessian matrix $H$ in this case. 
The smallest eigenvalue for two defects pairs symmetrically positioned along the real axis is always positive, 
which means that this formation is always stable in this triplet model.

\subsubsection*{Four defects}

Assume that $q_0=-1/2$ exists in the centre of triplets and three topological defects with charge $q_k=+1/2$, $k=1,2,3$, exist at symmetric positions. 
The coordinates of these defects are defined by $z_0=0$, $z_1=s$, $z_2=s\omega$, $z_3=s\omega^2$, $\omega =e^{2\pi \rmi/3}$, $s>0$. 
The energy that depends only on $z_k$ is given by 
\begin{align}
    F_d =-\frac{3}{4}\left[3\log\frac{a^3+2s^3}{(a^3 - s^3)^2} +\log(1-s^6)\right] - \frac{9}{4}\log\left(\frac{a^6-1}{a^3-a^2+1}\right) + \frac{5}{4}\log |g'(0)| - \frac{3}{4}\log 3. \nonumber 
\end{align}
The extrema of $F_d$ with respect to $s$ is obtained by
\begin{align}
    \frac{\partial F_d}{\partial s} = -\frac{3}{4}\left[\frac{18s^2}{a^3 + 2s^3} + \frac{18s^2}{a^3 - s^3} - \frac{6s^5}{1-s^6}\right] = -\frac{9}{2}s^2\left[\frac{3(2a^3 + s^3)}{(a^3 + 2s^3)(a^3 - s^3)} - \frac{s^3}{1-s^6} \right]=0,\nonumber 
\end{align}
which yields the following $9$-th-order polynomial formula with respect to $s$: 
\begin{align}
p_3(t) = t^3 + 7a^3t^2 - (3-a^6)t - 6a^3 = 0,\quad t = s^3.
\end{align}
Note that $p_3(0)=-6a^3<0$, $p_3(1)=a^6+a^3-2>0$, and 
\begin{align}
    p_3'(t)= 3t^2 + 14a^3t - 3+a^6,\quad p_3''(t) = 6t + 14a^3 > 0,
\end{align}
which means that $p_3(t)=0$ has a root in the range $0<t<1$.

The blue dotted line in Figure~\ref{fig:three_two} shows the smallest eigenvalue of the Hessian matrices for the three cases. 
When $\Delta > 1.481$, the smallest eigenvalue is positive, which means the formation of three topological defects becomes stable when $\Delta$ becomes larger than approximately $1.481$.

\section{Discussion and conclusion}\label{sec:5}

This paper provides analytical formulas for Frank free energy in the presence of topological defects based on analogies to vortex dynamics. 
The energy representation was obtained using the approach of Hamiltonians for vortex dynamics. 
The analytical formulas derived in this paper were used to obtain the positions of defects in doublet and triplet regions. 
The proposed formulas are explicit and give a more accurate condition for $\Delta$ when the transition of formation of topological defects occurs. 

Our analytical results closely match the experimental results reported by Ienaga {\em et al.}, particularly in the case of doublets~\cite{Ienaga2023-qm}. 
However, the defect distances $D_{++}$ obtained using the actual experiments conducted by Ienaga~{\it et al.} were found to be shorter than those obtained from our analytical methods. 
This suggests that the influence of cells must be considered, as we assumed cells to behave as nematic liquid crystals.

Although the present paper focuses on simply connected domains, 
it is important to note that Miyazako and Sakajo derived an analytical formula for cell alignments in doubly connected domains~\cite{Miyazako_undated-ue}. 
The derivation of the explicit formulas for energy in doubly connected domains is an ongoing work. 
Using the analytical energy formulas and the theory of quadrature domains in doubly connected domains described in~\cite{crowdy2020solving_book}, 
it might become possible to study ferromagnetic and antiferromagnetic order in bacterial vortex lattices in more complicated geometries, such as three merging circles with an inner holes. 

{\bf Data Accessibility:} This article has no additional data. 

{\bf Authors’ Contributions:} All authors contributed to the content of this paper.

{\bf Competing Interests:} We declare no competing interests. 

{\bf Funding:} The ﬁrst author is supported by JSPS KAKENHI Grant No. JP24KJ0041. The second author
is partially supported by JSPS KAKENHI, Grant No. JP23H00086 and JP23H04406.

{\bf Acknowledgements:} The ﬁrst author thanks Prof. Darren G. Crowdy at Imperial College London for
fruitful discussions.

\appendix

\section{Hessian}
We derive the Hessian with complex coordinates $z_k$ and $\conj{z_k}$. 
Because the derivatives with respect to $ z_k$ and $ \conj{z_k}$ are given by  
\begin{align}
    \frac{\partial}{\partial z_k} = \frac{1}{2}\left(\frac{\partial}{\partial x_k} - \rmi \frac{\partial}{\partial y_k} \right),\ \frac{\partial}{\partial \conj{z_k}} = \frac{1}{2}\left(\frac{\partial}{\partial x_k} + \rmi \frac{\partial}{\partial y_k} \right),
\end{align}
we have 
\begin{align}
    &\frac{\partial^2 }{\partial x_k \partial x_l}= \left(\frac{\partial}{\partial z_k} + \frac{\partial}{\partial \conj{z_k}} \right)\left(\frac{\partial}{\partial z_l} + \frac{\partial}{\partial \conj{z_l}} \right) = \frac{\partial^2}{\partial z_k \partial z_l} + \frac{\partial^2}{\partial \conj{z_k} \partial z_l} +\frac{\partial^2}{\partial z_k \partial \conj{z_l}} +\frac{\partial^2}{\partial \conj{z_k} \partial \conj{z_l}}, \\
    &\frac{\partial^2 }{\partial x_k \partial y_l}= \rmi \left(\frac{\partial}{\partial z_k} + \frac{\partial}{\partial \conj{z_k}} \right)\left(\frac{\partial}{\partial z_l} - \frac{\partial}{\partial \conj{z_l}} \right) = \rmi \left(\frac{\partial^2}{\partial z_k \partial z_l} + \frac{\partial^2}{\partial \conj{z_k} \partial z_l} -\frac{\partial^2}{\partial z_k \partial \conj{z_l}} -\frac{\partial^2}{\partial \conj{z_k} \partial \conj{z_l}}\right), \\
    &\frac{\partial^2 }{\partial y_k \partial y_l}= -\left(\frac{\partial}{\partial z_k} - \frac{\partial}{\partial \conj{z_k}} \right)\left(\frac{\partial}{\partial z_l} - \frac{\partial}{\partial \conj{z_l}} \right) = -\frac{\partial^2}{\partial z_k \partial z_l} + \frac{\partial^2}{\partial \conj{z_k} \partial z_l} +\frac{\partial^2}{\partial z_k \partial \conj{z_l}} - \frac{\partial^2}{\partial \conj{z_k} \partial \conj{z_l}}.
\end{align}
Using this transformation, the Hessian with respect to $x_k$ and $y_l$, $k,l=1,\ldots,N$, is given by
\begin{align}
    \begin{pmatrix}
        \dfrac{\partial^2}{\partial x_k\partial x_l} & \dfrac{\partial^2}{\partial x_k\partial y_l}\\
        \dfrac{\partial^2}{\partial y_k\partial x_l} & \dfrac{\partial^2}{\partial y_k\partial y_l}
    \end{pmatrix} = 
    \mathcal{J}\begin{pmatrix}
        \dfrac{\partial^2}{\partial z_k\partial z_l} & \dfrac{\partial^2}{\partial z_k\partial \conj{z_l}}\\
        \dfrac{\partial^2}{\partial \conj{z_k}\partial z_l} & \dfrac{\partial^2}{\partial \conj{z_k}\partial \conj{z_l}}
    \end{pmatrix}\mathcal{J}^\top,\quad \mathcal{J} \equiv  \begin{pmatrix}
        1 & 1 \\ \rmi & -\rmi
    \end{pmatrix}.
\end{align}

\bibliographystyle{unsrt}
\bibliography{reference2}

\end{document}